\documentclass[aps,prl,reprint,twocolumn,superscriptaddress,floatfix,nofootinbib,longbibliography]{revtex4-1}
\usepackage{amsthm}
\usepackage{amsmath,amssymb,color,comment,physics}
\usepackage[makeroom]{cancel}
\usepackage[caption=false]{subfig}
\usepackage{mathrsfs}
\usepackage{graphicx}
\usepackage{subfig}
\usepackage[countmax]{subfloat}
\usepackage[english]{babel}
\usepackage{dsfont}
\usepackage[bookmarks=true,colorlinks,linkcolor=OrangeRed,urlcolor=NavyBlue,citecolor=RoyalBlue]{hyperref}
\usepackage[dvipsnames]{xcolor}
\usepackage{braket}
\usepackage{mathtools}
\usepackage{soul} 

\definecolor{mygold}{rgb}{0.93,0.59,0.13}
\definecolor{mypurple}{rgb}{0.49,0.18,0.56}

\definecolor{philipp}{rgb}{1,.4,.3}

\definecolor{mygreen}{rgb}{0.25,0.5,0.25}

\begin{document}

\preprint{APS/123-QED}

\title{Emergent disorder and sub-ballistic dynamics in quantum simulations of the Ising model using Rydberg atom arrays}

\author{Ceren B.~Da\u{g}}
\email{ceren_dag@g.harvard.edu}
\affiliation{Department of Physics, Harvard University, 17 Oxford Street Cambridge, MA 02138, USA}
\affiliation{ITAMP, Harvard-Smithsonian Center for Astrophysics, Cambridge, Massachusetts, 02138, USA}

\author{Hanzhen Ma}
\affiliation{Department of Physics, Harvard University, 17 Oxford Street Cambridge, MA 02138, USA}

\author{P.~Myles Eugenio}
\affiliation{Department of Physics, University of Connecticut, Storrs, CT 06269, USA}
\affiliation{Department of Physics, Harvard University, 17 Oxford Street Cambridge, MA 02138, USA}

\author{Fang Fang}
\affiliation{QuEra Computing Inc., 1284 Soldiers Field Rd, Boston, MA 02135, USA}
\affiliation{Department of Physics, Harvard University, 17 Oxford Street Cambridge, MA 02138, USA}

\author{Susanne F.~Yelin}
\affiliation{Department of Physics, Harvard University, 17 Oxford Street Cambridge, MA 02138, USA}


\begin{abstract}

Rydberg atom arrays with Van der Waals interactions provide a controllable path to simulate the locally connected transverse-field Ising model (TFIM), a prototypical model in statistical mechanics. Remotely operating the publicly accessible Aquila Rydberg atom array, we experimentally investigate the physics of TFIM far from equilibrium and uncover significant deviations from the theoretical predictions. Rather than the expected ballistic spread of correlations, the Rydberg simulator exhibits a subballistic spread, along with a logarithmic scaling of entanglement entropy in time — all while the system mostly retains its initial magnetization. By modeling the atom motion, we trace these effects to an emergent disorder in Rydberg atom arrays, which we characterize with a minimal random spin model. We further experimentally explore the different dynamical regimes hosted in the system by varying the lattice spacing and the Rabi frequency. Our findings highlight the crucial role of atom motion in the many-body dynamics of Rydberg atom arrays at the TFIM limit, and propose simple benchmark measurements to test for its presence in future experiments.
\end{abstract}

\maketitle

\begin{figure}
\centering
\includegraphics[width=1.0\columnwidth]{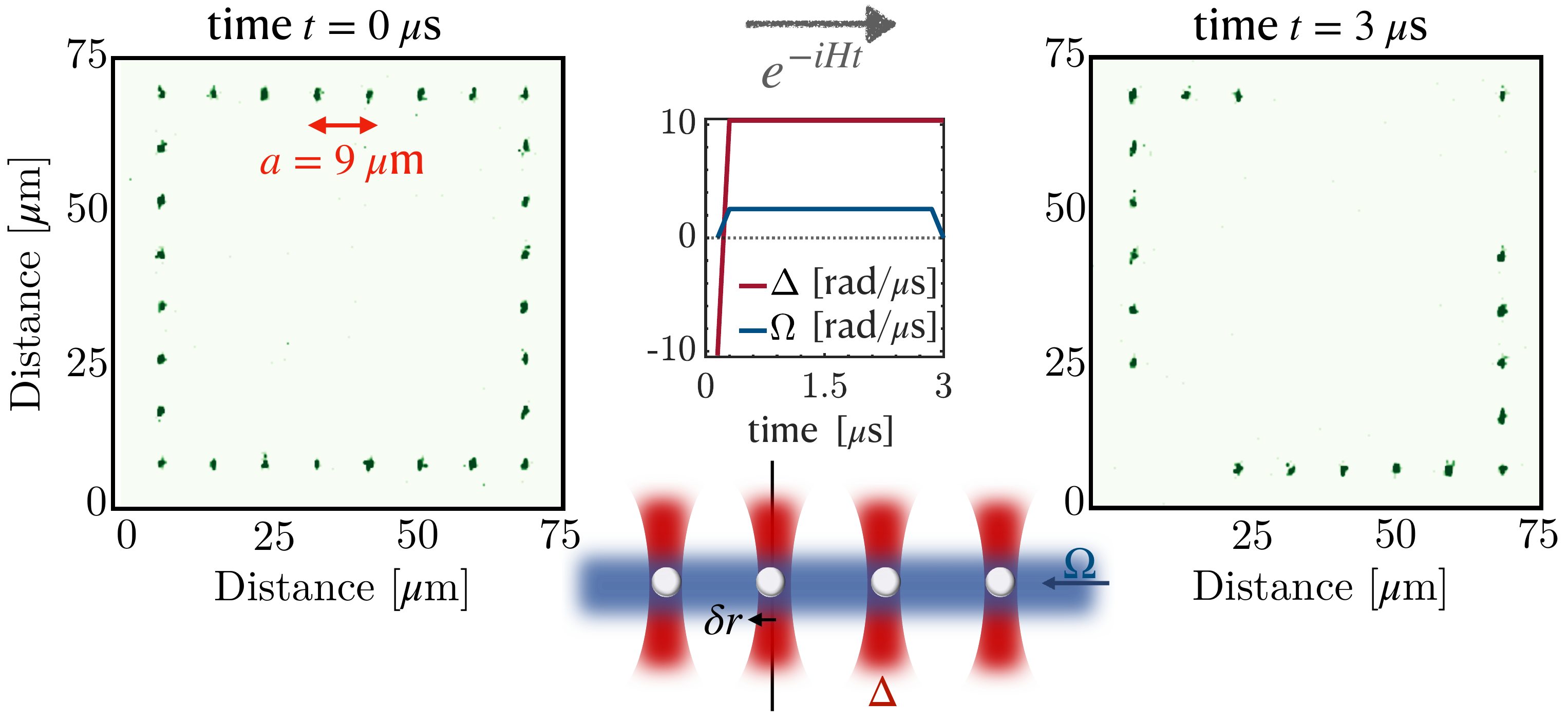}
\caption{A measurement sequence on Aquila atom array trapped in tweezers with all atoms in the ground state at $t=0$ evolving to the final state measured at $t=3\mu$s. The machine images, provided by QuEra Computing, shows our lattice geometry and the pulse sequence of the experiment where the Rabi frequency $\Omega$ and the detuning $\Delta$ are suddenly quenched, i.e.,~ramped in $50$ns. Due to thermal fluctuations, atoms move which gives rise to a positional uncertainty $\delta r$.}
\label{Fig0}
\end{figure}

\begin{figure}
\centering
\includegraphics[width=1.0\columnwidth]{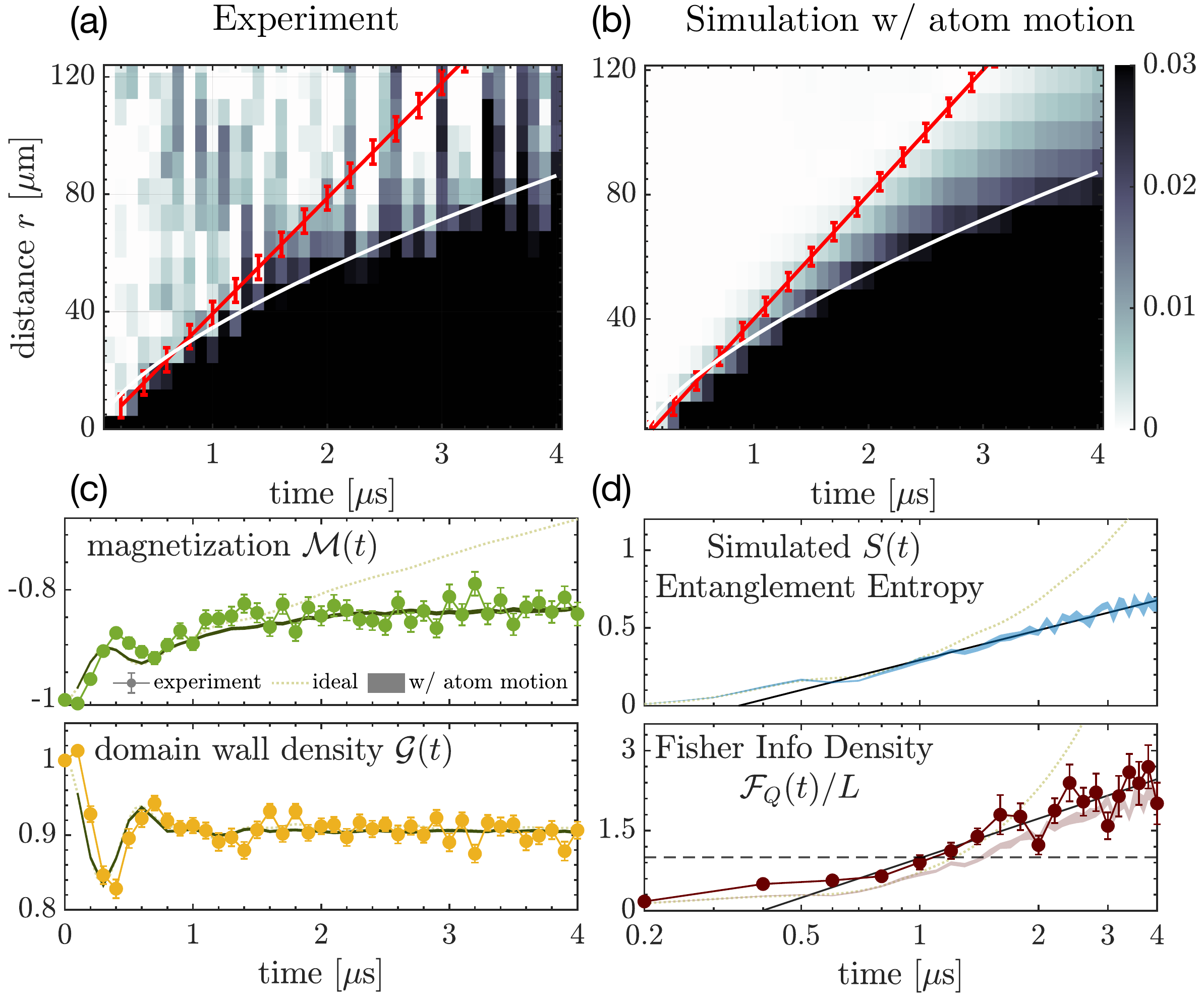}
\caption{(a) Measured and (b) simulated lightcone of quasi-particles generated following a quench to Rabi frequency $\Omega = 2.2 \text{ rad}/\mu$s at the TFIM limit and plotted with perceptually uniform colormaps. The quasi-particles slow down due to atom motion following a subballistic lightcone, solid-white, instead of ballistic, solid-red. (c) Measured (markers) and simulated (shades) magnetization $\mathcal{M}(t)$ and domain-wall density $\mathcal{G}(t)$ compared to the ideal simulation (dotted). (d) Simulated bi-partite entanglement entropy $S(t)$ (shade), and measured (marker) and simulated (shade) quantum Fisher information density $\mathcal{F}_Q(t)/L$, all exhibiting a logarithmic scaling in time when atom motion is modeled. The solid-black line is used to guide the line to show the logarithmic scaling. The ideal $S(t)$ and $\mathcal{F}_Q(t)/L$ are linear and power-law in time, respectively. All error bars are standard error of the mean (s.e.m).}
\label{Fig1}
\end{figure}

Unprecedented control of Rydberg atoms in tweezer traps present a unique opportunity to physicists to prepare fundamentally interesting and technologically useful quantum states of matter \cite{doi:10.1126/science.aax9743,doi:10.1126/science.aav9105,keesling2019quantum,browaeys2020many,scholl2021quantum,ebadi2021quantum,bernien2017probing,semeghini2021probing}. A particularly intriguing aspect of Rydberg atom arrays interacting with Van der Waals (vdW) interactions is the possibility of realizing the prototypical model of quantum statistical mechanics: the transverse field Ising model (TFIM). Although there are works studying Ising models on atom arrays \cite{labuhn2016tunable,PhysRevLett.120.180502,PhysRevA.110.053321}, the regime where the longitudinal field is kept at zero, i.e.,~TFIM regime, placing the setup at the blockade radius, was left uncharted. Besides exhibiting a quantum phase transition \cite{sachdev2001quantum}, TFIM also has rich behavior when driven out of equilibrium \cite{PhysRevA.69.053616,PhysRevLett.106.227203,Calabrese_2012,PhysRevLett.121.016801,PhysRevLett.123.115701,2020arXiv200412287D,PhysRevB.107.L121113}. One nonequilibrium behavior of TFIM is the ballistic spread of correlations when it is quenched, which originates from the constant propagation speed of the excitations generated due to the quench \cite{Calabrese_2012}. This also leads to a volume-law entangled state with a linearly increasing entanglement entropy in time \cite{Calabrese_Entropy}. Here we report on the experimental observation of instead a subballistic propagation in a one-dimensional Rydberg atom array under a quench [Figure \ref{Fig1}(a)], which subsequently generates sub-volume law entangled states with logarithmic increase of entanglement entropy in time against the theoretical prediction. 
This behavior does not originate from the constrained kinematics due to Rydberg blockade \cite{ebadi2021quantum}, because at the TFIM limit we are never within the blockade radius, which casts the TFIM limit as a different regime of Rydberg atom arrays from the PXP model limit \cite{turner2018weak}. Our tensor network simulations instead, trace this unexpected phenomenon to \textit{the motion of atoms} originating from the thermal fluctuations, Fig.~\ref{Fig0}. The atom motion, at $T\sim 15 \mu \text{K}$ temperature \cite{wurtz2023aquilaqueras256qubitneutralatom}, gives rise to emergent disorder, breaking the $Z_2$ symmetry in the TFIM. We also detect localized and thermal regimes in which the initial magnetization is either retained or lost, accompanied by a logarithmic or fast increase in quantum Fisher information (QFI), respectively. Motivated by these observations, we introduce a random spin model to describe the physics of the Rydberg atom array when it is mapped to TFIM, which can be used in future theoretical studies.

$Z_2$ symmetry breaking in anti-ferromagnetic TFIM, due to atom motion, prevents the formation of spontaneous ferromagnetism at the top of the spectrum, $ \ket{\psi_{\rm FM}}_{\pm}  \propto   \ket{\uparrow \uparrow \cdots \uparrow} \pm  \ket{\downarrow \downarrow \cdots \downarrow}$. Hence, the prevalent idea of utilizing sudden quenches, instead of slow ramps, to probe spontaneous symmetry breaking (SSB) phase transition in Ising models \cite{PhysRevLett.115.245301,PhysRevLett.121.016801,PhysRevLett.123.115701,PhysRevB.107.094432} is currently not attainable in Rydberg simulators due to significant atom motion in many-body dynamics. Nonetheless, we find a trace of the SSB transition of TFIM in the time-averaged magnetization and domain-wall density, with $Z_2$ symmetry explicitly broken. Our work demonstrates that operating Rydberg atom arrays at the blockade radius requires a treatment of the atom motion induced by the thermal fluctuations, and theoretically characterizes this particular regime of atom arrays with a random spin model.

\textit{Experimental setup.~}We remotely operate Aquila Rydberg atom array designed by QuEra Computing \cite{wurtz2023aquilaqueras256qubitneutralatom}. Figure \ref{Fig0} shows the rectangular ring geometry of our setup with $9\mu\text{m}$ distance between 28 atoms, and the quench protocol. Periodic boundary conditions enhance data quality by providing protection against edge effects and enabling faster statistics in many-body physics. The technical details and limitations of the setup can be found in Aquila whitepaper \cite{wurtz2023aquilaqueras256qubitneutralatom}. The Rydberg simulator Hamiltonian can be mapped to the spin$-1/2$ mixed-field Ising model with vdW interactions \cite{browaeys2020many,labuhn2016tunable,PhysRevLett.120.180502},
\begin{eqnarray}
H(t) &=& J \sum_r^L \sigma^z_r \sigma^z_{r+1} + \frac{J}{2^6}\sum_r^L \sigma^z_r \sigma^z_{r+2} \notag \\
&+& \sum_r^L h_x(t) \sigma^x_r -  \sum_r^L h_z(t)\sigma^z_r, \label{eq1}
\end{eqnarray}
with periodic boundary conditions and system size $L$.  The interaction strength, transverse and longitudinal fields read
\begin{eqnarray}
J &=& \frac{C_6}{4a^6}, h_x(t) = \frac{\Omega(t)}{2}, h_z(t)= \frac{1}{2}(\Delta(t)-\Delta_{\rm ising}),
\end{eqnarray}
in terms of the experimentally tunable parameters Rabi frequency $\Omega(t)$, detuning $\Delta(t)$, the atom distance $a$, and the Rydberg interaction coefficient $C_6=5.42\times 10^{-24} \textrm{rad Hz m}^6$. We define the TFIM limit to be the Rydberg Hamiltonian with detuning $\Delta(t)=\Delta_{\rm ising}$, where $\Delta_{\rm ising} = 4J \left(1+\frac{1}{2^6}\right)$. The latter expression guarantees $h_z(t)=0$ in Eq.~\eqref{eq1} reducing it to the TFIM, hence the name. We perform the pulse sequence in Fig.~\ref{Fig0} to implement the quench protocol. This corresponds to time evolving the spin polarized initial state~$\ket{\downarrow \downarrow \cdots \downarrow}$, i.e.,~all atoms are in the ground state with $\Delta(t=0) \ll 0$ and $\Omega(t=0)=0$, at the TFIM limit which is achieved with~$\Delta(t>50\text{ns})=10.34 \text{ rad}/\mu\text{s}$ and $\Omega(t>50\text{ns}) \leq 15.7 \text{ rad}/\mu\text{s}$. In other words, we change the Rabi frequency and detuning simultaneously in $50$ns from the initial state, and hold the system until the measurement time $t\leq 4 \mu\text{s}$. We compute the averaged connected equal-time correlators $\mathcal{G}_{r}^{\rm{conn}}(t) = \frac{1}{L} \sum_i G_{i,r}^{\rm{conn}}(t)$ where $  G_{i,r}^{\rm{conn}}(t) = \langle \sigma_i^z(t)\sigma_{i+r}^z(t)\rangle - \langle \sigma_i^z(t)\rangle \langle \sigma_{i+r}^z(t)\rangle $  in order to determine the lightcone dynamics, in addition to the time-evolved magnetization $\mathcal{M}(t) \coloneq  \frac{1}{L} \sum_i  \braket{\sigma_i^z(t)}$ and domain-wall density $\mathcal{G}(t) \coloneq \frac{1}{L} \sum_i^L \braket{\sigma^z_i(t) \sigma^z_{i+1} (t)}$ to probe the $Z_2$ symmetry breaking and the dynamical regimes. We compute the fluctuations of normalized total magnetization $\frac{L}{2}\mathbb{M}(t) \coloneq \frac{1}{2} \sum_i  \sigma_i^z(t) $ with $\mathcal{F}_Q(t)/L = L \left( \braket{\mathbb{M}(t)^2}-\braket{\mathbb{M}(t)}^2 \right)$, which is also the QFI of the generated dynamical pure state in the unitary numerical simulations, and indirectly probes the entanglement \cite{Smith_2016}. When we compute $ \mathcal{F}_Q(t)/L$ for the experimental data, it acts as an upper bound to the actual QFI of the dynamical state \cite{PhysRevLett.72.3439}.

\textit{Subballistic spread of correlations and entanglement dynamics.~}Since the interactions are anti-ferromagnetic in the Rydberg atom array, a quench from $\ket{\downarrow \downarrow \cdots \downarrow}$ excites the many-body system to the high energy states. In ideal conditions, this quench protocol would give rise to a linear lightcone probed by $G_{i,r}^{\rm conn}$ at any $i$ due to translational invariance as computationally confirmed with time-evolving block decimation (TEBD) algorithm in tensor networks \cite{ITensor}, see SM \cite{supp}. Consequently the entanglement entropy of the dynamical state increases linearly in time, black-solid in Figure \ref{Fig1}(d), a signature of the entangled pairs of quasi-particles generated due to quench \cite{Calabrese_Entropy}. In this dynamical regime, $\mathcal{M}$ exponentially decays to zero in time \cite{Calabrese_2012}, whereas $\mathcal{G}$ remains larger than $1/2$ in long-time limit \cite{PhysRevLett.123.115701}, Figure \ref{Fig1}(c) showing that the initial magnetization is eventually lost, and the dynamically generated state always has a finite domain-wall density. 

The lightcone $\mathcal{G}^{\rm conn}_r(t)$ computed with the experimental data in Figure \ref{Fig1}(a) after readout-error mitigation \cite{supp} demonstrates a subballistic spread of correlations, instead of ballistic, for a quench from polarized state to $h_x/J=0.43$ (or $\Omega \sim 2.2\text{ rad}/\mu$s at $a=9\mu$m). The legend maximum is set to the experimental resolution \cite{supp}. The nonzero values outside of the lightcone is expectantly bounded by this resolution. To determine the origin of this behavior in the many-body dynamics, we simulate the Rydberg atom array Hamiltonian on tensor networks \cite{ITensor} and model the error sources including atom motion, spatial detuning fluctuations, shot-to-shot fluctuations in the detuning and Rabi frequency. For details on the modeling of atom motion and other uncertainties, see End Matter and SM \cite{supp}. We find that the dominant mechanism underlying our experimental observations is the atom motion. The tensor network simulation given in Figure \ref{Fig1}(b) reproduces the subballistic behavior well, [see \cite{supp} for temporal cross-sections with statistical error bars]. Ideally the quasi-particles would exhibit a ballistic light-cone with a maximum speed of $v_c \sim 40$m/s which can be analytically determined from $v_c\simeq 2\Omega a$ if $h_x/J<1$\cite{Calabrese_2012}, shown with green and red in Figs.~\ref{Fig1}(a) and (b), respectively. However, the quantum many-body dynamics slow down under significant atom motion turning the spread of correlations to subballistic, with $r \simeq 34.6 t^{0.66}$, from ballistic. This is successfully captured by the numerical modeling in Fig.~\ref{Fig1}(b) with $r \simeq 34.5 t^{0.67}$. 
Hence, we find that the simple theoretical picture painted above breaks down on a Rydberg simulator due to atom motion. Let us note that other uncertainties do not affect the ballistic behavior of the clean TFIM \cite{supp}. For this set of parameters, we are right at the blockade radius, where effects from atom-phonon coupling are in fact predicted for the vibrational ground and the first excited Fock states \cite{PhysRevLett.132.133401}. However such low-energy quantum vibrational fields are not an accurate description for a typical atom array setup \cite{supp}. From another perspective, given that a decohering TFIM would lead to halting spread of correlations over time $v_c=0$, subballistic spread is, in fact, a plausible trend in the intermediate timescales.

Other observables follow suit. The numerically computed entanglement entropy exhibits a sub-volume law behavior with a logarithmic increase in time, Fig.~\ref{Fig1}(d). 
Simulated QFI in Fig.~\ref{Fig1}(d) for unitary dynamics also exhibits a logarithmic increase in time, and importantly it successfully models the experimental $\mathcal{F}_Q(t)$ and its temporal scaling. An array with no atom motion would exhibit a power-law increase of QFI in time, dotted line $\mathcal{F}_Q(t)/L \propto t^{1.6}$. Hence the data demonstrates the role of atom motion in the entanglement of the states generated far from equilibrium. Remembering that the particular $\mathcal{F}_Q(t)$ we study here is a cumulative effect of all possible spatial correlations in the $z-$basis, the logarithmic scaling is consistent with the subballistic lightcone in Fig.~\ref{Fig1}(a).
Furthermore, the magnetization decays to a finite value, Fig.~\ref{Fig1}(c) implying that the many-body system partially retains its initial condition. The domain-wall density does not differ significantly from the ideal case. This is because $\mathcal{G}(t)$ cannot differentiate between explicit and spontaneous symmetry breaking.

\textit{Minimal random spin model.~}Inspired by these experimental observations and to gain more insight about the role of atom motion in many-body dynamics, we propose a minimal yet realistic many-body model for the TFIM limit of the Rydberg atom arrays. This time-independent minimal model (MM) incorporates only the effect of atom motion into the Hamiltonian with the atom positional uncertainty $\delta r$,
\begin{eqnarray}
    H^{\rm eff} &=& \sum_r J_r  \sigma^z_r \sigma^z_{r+1} +  \frac{\Omega}{2} \sum_r\sigma^x_r -  \sum_r h_r \sigma^z_r, \label{eq:emergentSpin}\\
    J_r &=& \mathcal{N}\left( \frac{1}{4}\frac{C_6}{a^6},\frac{3}{2}\frac{C_6 \delta r}{a^7} \right), h_r = \mathcal{N}\left(0, 3\frac{C_6 \delta r}{a^7}\right),
\end{eqnarray}
where $\mathcal{N}(\mu_N,\sigma_N)$ stands for normal distribution with mean $\mu_N$ and standard deviation $\sigma_N$. Time evolving Eq.~\eqref{eq:emergentSpin} results in qualitative agreement with the numerical simulation of Rydberg atom array Hamiltonian, where atom motion is modeled both by random initial atom positions and velocities together with other uncertainties \cite{supp}. Hence, MM provides a simpler alternative to computationally expensive modeling at the TFIM limit of Rydberg atom arrays. In MM, atom motion effectively introduces random disorder to the Hamiltonian $\mathcal{W} \equiv \sigma_N(h_r)/\mu_N(J_r) \sim 12 \delta r/a$, where $\sigma_N(h_r)$ and $\mu_N(J_r)$ represent the standard deviation of $h_r$ and the mean of $J_r$. Although such a disorder is always present in Rydberg simulators, its effect is suppressed and is likely to become insignificant in many-body dynamics away from the TFIM limit where $\mu_N(h_r) \gg \sigma_N(h_r)$ holds. 

\begin{figure}
\centering
\includegraphics[width=1.0\columnwidth]{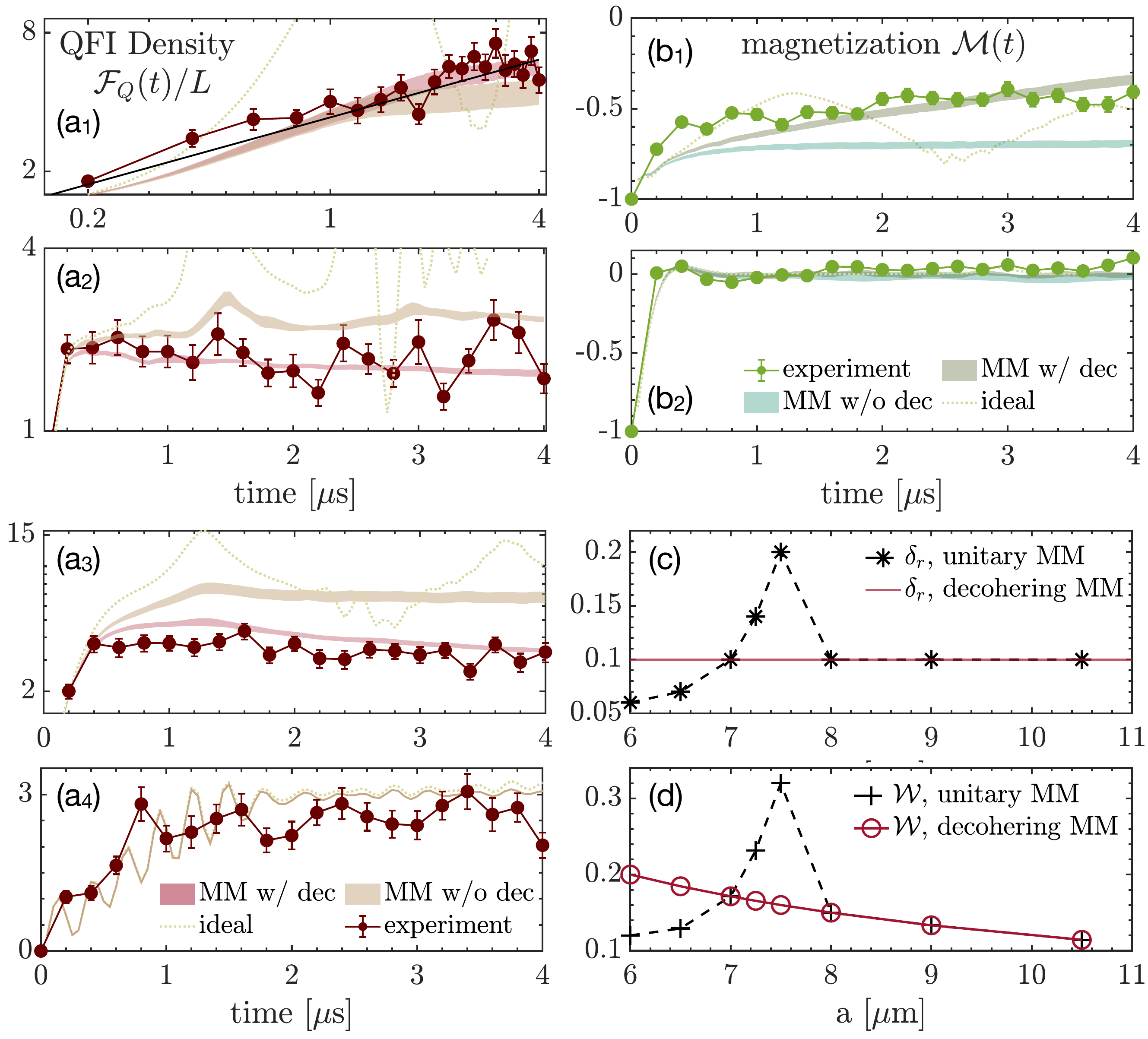}
\caption{Naturally disordered quantum many-body system following a quench to $\Omega=11.7$ rad$/\mu$s at the TFIM limit numerically (shades) and experimentally (markers) probed with (a) Quantum Fisher information (QFI) density and (b) magnetization for different lattice constants. Panels (a): (a$_1$), (a$_2$), (a$_3$) and (a$_4$) respectively plot the QFI density for $a=7\mu$m, $a=8\mu$m, $a=7.5\mu$m and $a=10.5\mu$m. Panels (b): (b$_1$) and (b$_2$) respectively plot the magnetization for $a=7\mu$m and $a=8\mu$m. For (a) and (b), the legends include the results of the experiment, ideal setup with no atom motion (dotted), the minimal model (MM) in Eq.~\eqref{eq:emergentSpin} modeling the atom motion in the presence and absence of decoherence. Fig.~(a$_1$) is plotted in semi-log scale to show the logarithmic increase of QFI density in time. (c) The positional uncertainty $\delta r$ for different lattice constants in the unitary and decohering MM 
and (d) the resulting disorder strength $\mathcal{W}$. All error bars are s.e.m. Error bars for the MM are due to 10 different statistically similar systems with different configurations of $J_r$ and $h_r$.}
\label{Fig2}
\end{figure}

Next we experimentally test the dependence of disorder  $\mathcal{W}$ on the lattice spacing $a$. For this, we quench to $\Omega\sim 11.7 \text{ rad}/\mu$s at the TFIM limit and vary only the lattice constant between $a=6-10.5\mu$m with the ratio $a/R_b=1-1.21$ when $a$ increases where $R_b=\left(C_6/\sqrt{\Omega^2+\Delta^2}\right)^{1/6}$ is the Rydberg radius \cite{wurtz2023aquilaqueras256qubitneutralatom}. Fig.~\ref{Fig2}(a) panels show the QFI for select lattice constants. We spot a large difference between experiment and ideal simulations due to atom motion, except at $a=10.5\mu$m, (a$_4$). At large atom distances, QFI slowly increases to its maximum value with a good match between simulations in the presence and absence of atom motion. This suggests that the system approaches its clean limit $W\sim 0$, as the lattice constant increases, $h_x/J \gg 1$. At $a=7\mu$m corresponding to $h_x/J < 1$, plotted in (a$_1$), the temporal scaling of QFI is consistent with a logarithmic scaling surpassing $\mathcal{F}_Q/L = 1$, and hence 
suggesting  many-body localization-like behavior \cite{2006AnPhy.321.1126B,Nandkishore_2015,Smith_2016}. Consistently, the initial magnetization is mostly retained in the experiment, Fig.~\ref{Fig2}(b$_1$). In contrast, at $a=8\mu$m corresponding to $h_x/J > 1$ plotted in (a$_2$), QFI quickly increases to its maximum value in $0.2\mu$s and remains constant after both in numerics and experiment, while the initial magnetization quickly decays to zero, (b$_2$). Hence, we bear witness to localized and delocalized regimes in the quench dynamics of Rydberg atom arrays due to atom motion. 

The positional uncertainty is kept fixed at $\delta r=0.1\mu$m in the simulations, Fig.~\ref{Fig2}, as this value is set by the trap width. We notice that MM does not capture the experimental data in all lattice constants, see $a=7.5\mu$m in (a$_3$) and \cite{supp}. It is curious to note that these lattice distances correspond to $h_x/J$ which are closest to the quantum critical point (QCP) in the ground and the most-excited states. 
We develop two approaches to address this mismatch: (i) We treat $\delta r$ in MM as a fitting parameter and determine the profile of $\delta r(a)$, black-dashed Fig.~\ref{Fig2}(c), by fitting $\mathcal{F}_Q/L$ of MM to those of experimental data \cite{supp}. This results in a peak in the disorder profile in Fig.~\ref{Fig2}(d), which is, in fact, a signature of physics that extends beyond the capability of a unitary model \cite{fang2024probingcriticalphenomenaopen}. (ii) We keep $\delta_r=0.1\mu$m and introduce classical Markov noise to our simulations to mimic the decoherence process in $z-$basis \cite{PhysRevLett.128.070402,Kampen1992,PhysRevLett.118.140403} with $\gamma_r(t)=\mathcal{N}\left(0, 0.6\text{MHz}\right)$ in $h_r \rightarrow h_r + \gamma_r(t)$ (red and gray shades in (a) and (b)). Physically, such a process originates from laser noise and intermediate state scattering \cite{fang2024probingcriticalphenomenaopen,PhysRevA.97.053803,kozlej2025adiabaticstatepreparationthermalization}. Decohering MM results match the experimental data well and hence demonstrate the role of decoherence besides atom motion. With decoherence included into the modeling, the disorder profile smoothly decays in lattice spacing, Fig.~\ref{Fig2}(d). In conclusion, both versions of MM with the experimentally determined disorder profiles can be utilized to further explore the TFIM limit of Rydberg atom arrays \cite{supp}. 

\begin{figure}
\centering
\includegraphics[width=1.0\columnwidth]{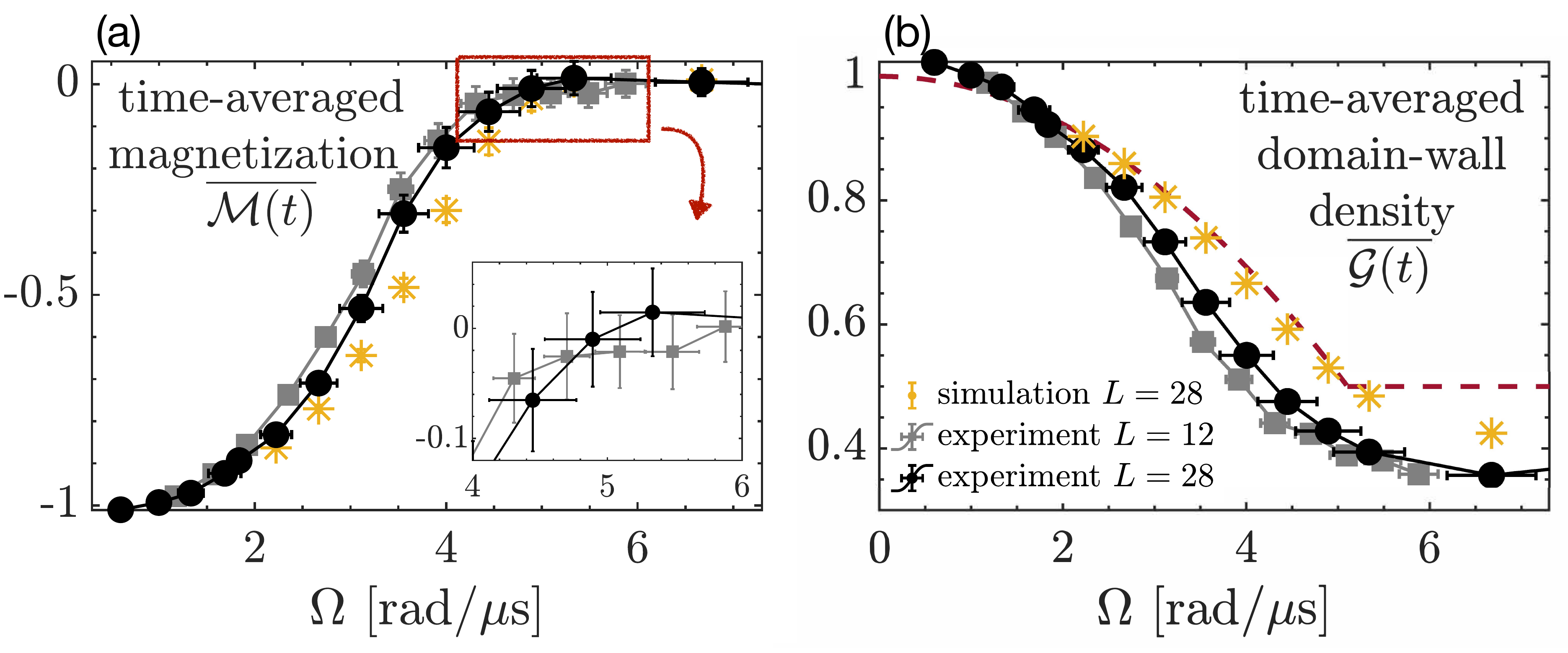}
\caption{Time-averaged (a) magnetization and (b) domain-wall density to experimentally probe a trace of $Z_2$ symmetry breaking of spontaneous symmetry breaking transition at the clean TFIM. Experiments are performed with quenches to different Rabi frequencies $\Omega$ and hence transverse field $h_x$ at the TFIM limit at fixed lattice constant $a=9\mu$m. Orange-stars are the simulated results with all uncertainties included. Inset in (a) zooms on the crossover region. (b) also shows the ideal TFIM results in the infinite time limit with red-dashed line. All error bars are s.e.m.}
\label{Fig3}
\end{figure}

\textit{Trace of SSB transition.~}Ferromagnetic TFIM with vdW interactions, $h_z=0$, hosts an SSB transition between an SSB ferromagnet and a paramagnet at the QCP $h_{\rm qcp}=1.027J$. Meanwhile $\mathcal{G}(t)$ in the thermodynamic limit and infinite time exhibits a discontinuity at the SSB transition of ferromagnetic TFIM signalling the transition in the wake of a quench \cite{PhysRevLett.123.115701}. The same physics holds in the anti-ferromagnetic TFIM: Following a quench from a polarized state to any $h_x$, time-averaged magnetization is $\overline{|\mathcal{M}(t\rightarrow \infty)|}=0$. Whereas $\overline{\mathcal{G}(t\rightarrow \infty)} > 1/2$ and $\overline{\mathcal{G}(t\rightarrow \infty)}=1/2$ for a quench to $h_x<h_{\rm qcp}$ and $h_x \geq h_{\rm qcp}$, respectively, as depicted with red-dashed lines in Fig.~\ref{Fig3}(b).

The disorder already breaks the $Z_2$ symmetry, leading to finite magnetization in time when we quench to $h_x < h_{\rm qcp}$, as in Fig.~\ref{Fig1}(d), instead of an exponential decay to zero \cite{Calabrese_2012}. We use this signature to define 
an order parameter-like quantity based on the time-average of experimental magnetization $\overline{|\mathcal{M}(t)|}$ and domain-wall density $\overline{\mathcal{G}(t)}$ between $[0,4]\mu$s, and plot these values in Figure \ref{Fig3} with respect to Rabi frequency for two different system sizes, $L=28$ and $L=12$, at fixed $a=9\mu$m, and hence at a fixed $\mathcal{W} \sim 0.13$. The magnetization exhibits an experimental crossover at $\Omega_c \sim 5$ rad$/\mu$s (with corresponding crossover transverse-field $h_c\sim J$) between two system sizes where $\overline{|\mathcal{M}(t)|} > 0$ is observed for $h_x < h_c$, and $\overline{|\mathcal{M}(t)|} \sim 0$ for $h_x \geq h_c$, see inset in Fig.~\ref{Fig3}(a). $\overline{|\mathcal{M}(t)|}$ tends to increase (decrease) with increasing system size in $h_x < h_c$ ($h_x > h_c$) with $h_c \sim h_{\rm qcp}$. The crossover trend of $\overline{|\mathcal{M}(t)|}$ is absent without the atom motion, even when the time-averaging is performed in the experimental time interval $[0,4]\mu$s \cite{supp}. Meanwhile, the domain-wall density reaches its minimum around $h_c$ with a value $\overline{\mathcal{G}(t)} \sim 0.4$, which undershoots $\frac{1}{2}$ due to explicit $Z_2$ symmetry breaking.
These results show that it is possible to take on a trace of $Z_2$ SSB transition in an experimental data of a disordered many-body system pushed out of equilibrium.

\textit{Conclusions and Outlook.~}This work experimentally demonstrates the remarkable effect of atom motion in quantum many-body dynamics on a Rydberg atom array at its blockade radius and provides a minimal model that captures the experimental observations. The minimal model encodes the atom motion as disorder in Ising interactions and longitudinal field, which differs from the random Ising models in the literature \cite{PhysRevB.88.014206,PhysRevLett.113.107204,PhysRevB.104.115159}. Hence we show that Rydberg atom arrays could be used to study the physics of disordered many-body systems without the need for local detuning control. This work also demonstrates the subtle interplay between atom motion and decoherence processes by clearly differentiating their contribution to the many-body dynamics. We further find that the decoherence effects are enhanced when the Hamiltonian is close to a quantum critical point.

TFIM limit is a fine-tuned point on a Rydberg atom array, which explains the importance of atom motion at this limit. Our results imply, more generally, that the fine-tuned points in other platforms \cite{RevModPhys.93.025001,richerme2014non,guo2024site}, can be susceptible to particle motion. While we exclusively discussed ferromagnetic initial states, the atom motion also dominates for quenches from Neel states, and possibly for slow ramps, to the low-energy sector of the Ising Hamiltonian. This suggests that more advanced cooling techniques are needed to realize clean Ising models with only transverse field on Rydberg systems. It may be interesting to consider whether the atoms can be synchronized \cite{samoylova2015synchronization} to combat the disorder. 
Quantizing the atom motion and considering the full atom-phonon Hamiltonian would take into account the backaction of the Rydberg interactions on the atomic motion, and hence can be an interesting future direction for determining the role of the atom-phonon entanglement in the many-body dynamics of Rydberg atom arrays. Our work characterizes a naturally disordered limit of Rydberg atom arrays with simple benchmarking experiments far from equilibrium, and demonstrates how to harness atom motion as a resource for exploring the physics of disordered many-body systems.

\textit{Acknowledgments.}~Authors thank A. Bylinskii, M. Kornjaca, T. Macrì, C.~W. Wächtler, Yuxin Wang, N. Yao and M.~D. Lukin for stimulating discussions, Amazon Web Services for the Cloud Credit for Research Program credits granted towards utilizing Aquila in this project, and QuEra Computing for technical assistance on Aquila, Majd Hamdan and P. Lopes for the machine images in Figure~\ref{Fig1}. C.B.D was supported with the ITAMP grant No.~2116679. SFY acknowledges support from the NSF through grant PHY-2207972 and the Q-Ideas HDR (OAC-2118310).

\begin{figure}
\centering
\includegraphics[width=1.0\columnwidth]{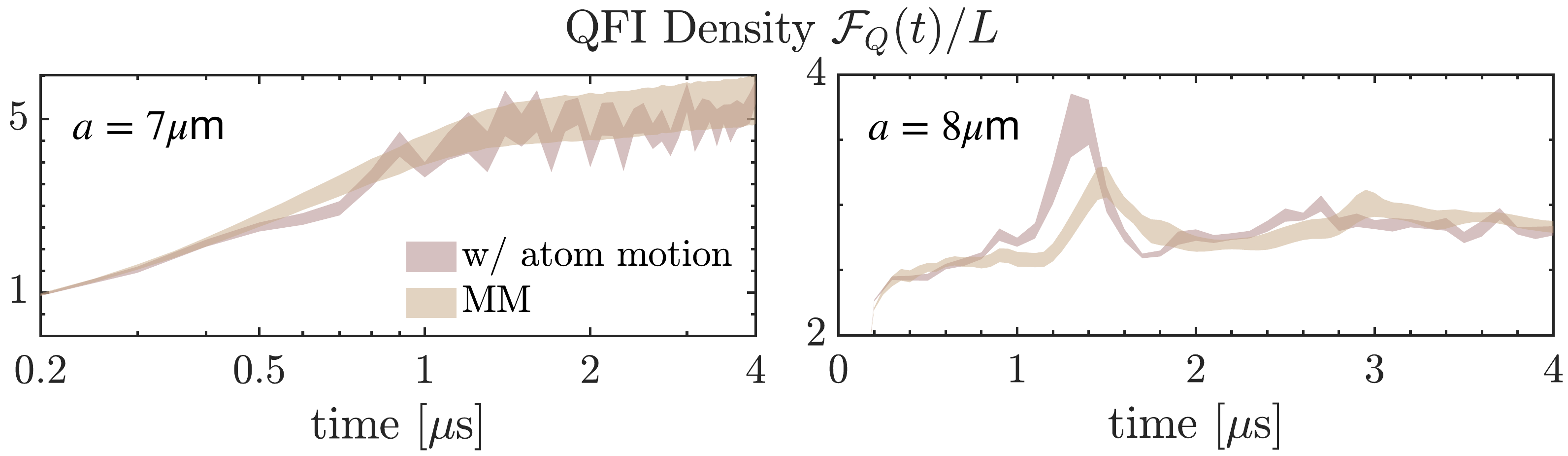}
\caption{Comparison of the QFI results from tensor network simulations of Rydberg atom array with atom motion treated classically and the time-evolution of minimal model Eq.~\ref{eq:emergentSpin} where $\delta_r=0.1\mu$m for lattice spacing $a=7\mu$m (left panel) and $a=8\mu$m (right panel) at $\Omega=11.7$ rad/$\mu$s.}
\label{FigEndMatter1}
\end{figure}

\textit{End Matter.}~Atom motion is modeled with a normal distribution for the uncertainty in the atom positions $\delta r=0.1\mu$m determining the initial positions of atoms, and Maxwell-Boltzmann distribution for the initial velocities determined by the atom temperature 15$\mu$K. The atom positions at later times are deterministically found by solving the equations of motion for each atom independently. These classically determined positions at each TEBD timestep set the geometry and modify the many-body Hamiltonian, which then affects the tensor network that time-evolves the quantum many-body state. This computationally intensive model can be approximated with our minimal model where atom motion is treated as emergent random disorder, see Fig.~\ref{FigEndMatter1}.

We calibrate our measurements before any many-body measurement by doing a two-photon resonance experiment on a collection of independent atoms on 2D geometry. In this experiment, we suddenly quench the Rabi frequency from 0 to the desired value, $\Omega$, keep the detuning constant at a value $\Delta$ and make a measurement at time $t_{\pi}=\pi/\Omega$. We repeat the same measurement for different detuning values obtaining the experimental excitation probability with respect to detuning. The excitation probability can be analytically derived via quantum optics techniques \cite{meystre2007elements}, $P_e(\Delta) = \frac{\Omega^2}{\Omega^2+\Delta^2} \sin^2\left(\sqrt{\Omega^2+\Delta^2} \frac{t_{\pi}}{2}\right)$.
We fit our experimental data to $P_e(\Delta)$ to find the systematic global detuning shift and the calibrated Rabi frequency. All reported values in the Letter are calibrated values. We then correct for the systematic detuning shift in all many-body measurements done in the same day. Since TFIM limit is a fine-tuned point in detuning, it is crucial to correct for this systematic uncertainty rather than modeling it in numerics. 
Calibration experiment also provides us 1-$\sigma$ deviation from the measured values. We apply bootstrap technique \cite{bootstrap_ana} to first observe the emergence of normal distribution for calibrated Rabi frequency and detuning, and then to extract 1-$\sigma$ deviation. The latter is used in numerical modeling of Rabi frequency and detuning fluctuations from shot-to-shot. We also locally apply bootstrap technique to each atom to determine the spatial detuning profile across the many-body chain, and find that this is roughly sinusoidal across the array \cite{supp}. Based on this method, we find negligible differences in the Rabi frequency of the atoms, and hence we do not include spatial variations in Rabi frequency in our numerical modeling. Finally, we apply error mitigation to the many-body data with confusion matrices derived for one- and two-body observables incorporating the readout errors \cite{PhysRevA.103.042605,supp}. In addition to that, we postselect the final bit strings based on the set initial state. 

\bibliographystyle{apsrev4-1}

%

\onecolumngrid
\newpage

\setcounter{equation}{0}
\setcounter{figure}{0}
\setcounter{table}{0}
\setcounter{page}{1}
\makeatletter
\renewcommand{\theequation}{S\arabic{equation}}
\renewcommand{\thefigure}{S\arabic{figure}}
\renewcommand{\bibnumfmt}[1]{[S#1]}
\setcounter{secnumdepth}{1}
\setcounter{secnumdepth}{2}

\begin{center}
\textbf{\Large Supplementary Material: Emergent disorder and sub-ballistic dynamics in quantum simulations of the Ising model using Rydberg atom arrays}
\end{center}
\hspace{5mm}
\begin{center}
{\large Ceren B.~Da\u{g}, Hanzhen Ma, P. Myles Eugenio, Fang Fang, Susanne F. Yelin}
\end{center}

\vspace{5mm}

\widetext


\section{\label{sec:1}Temporal cross-sections on light cone}

\begin{figure}
\centering
\includegraphics[width=1.0\columnwidth]{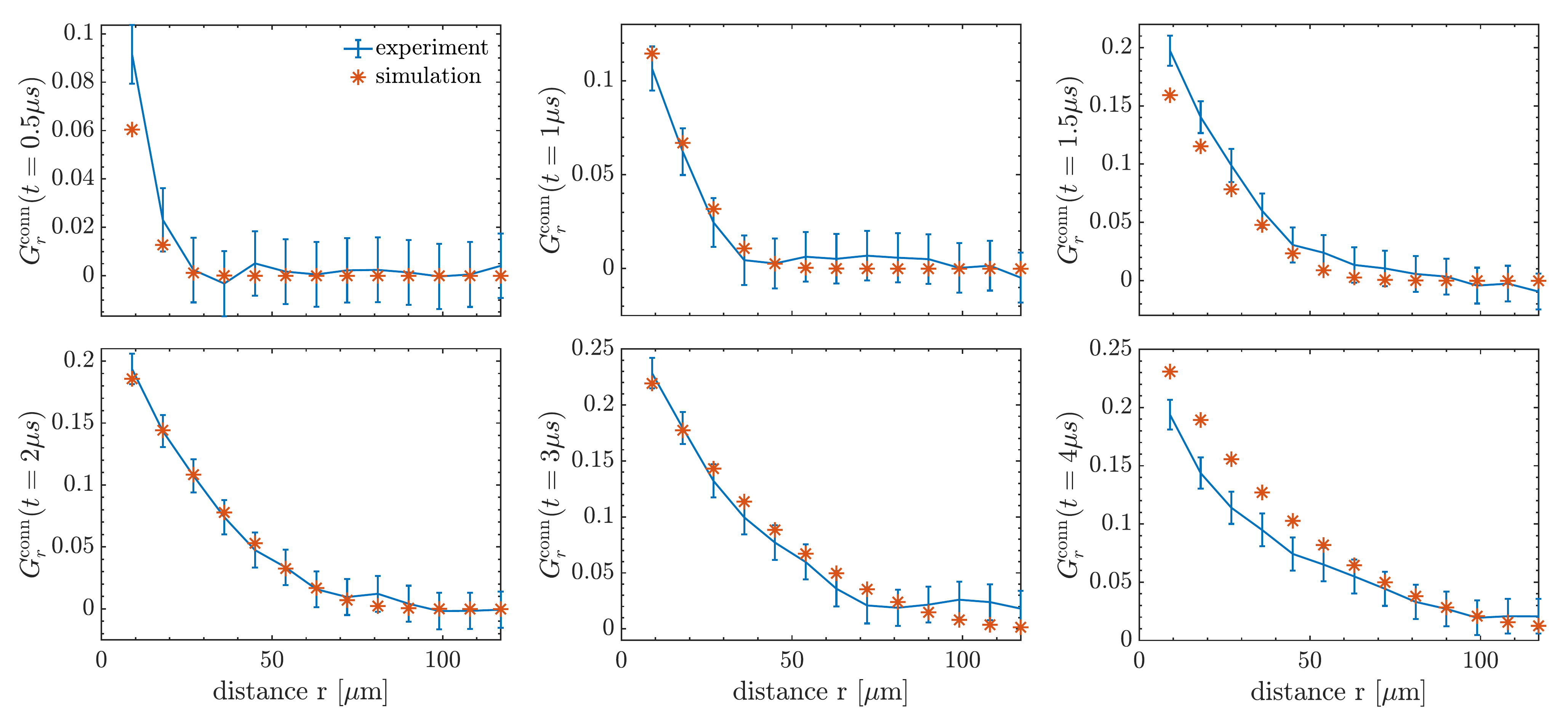}
\caption{Temporal cross-sections of $\mathcal{G}^{\rm conn}_r(t)$ at different times between $t=0.5\mu$s and $t=4\mu$s for the light cone plotted in main text, Figure 2(a). Experiment (solid-blue) and simulation with atom motion (orange-stars) match well.}
\label{SMFig1}
\end{figure}

In most of the many-body experiments, we collect 200 bit strings at each time. Due to post-selection based on the correct initial state, the number of useful bit strings reduces to around $\sim 160$ with a high enough post-selection rate. Hence the sampling number affects the standard error of the mean (s.e.m). The typical s.e.m. in the light cone figures is on the order of $\sim 0.03$, see Figs.~\ref{SMFig1}. This is the reason why we set the colorbar in the light cone figures to $0.03$, which is our experimental resolution. The numerical modeling match the experimental data well within the error bars. 

\section{\label{sec:4}Spatial detuning fluctuations}

\begin{figure}
\centering
\includegraphics[width=0.5\columnwidth]{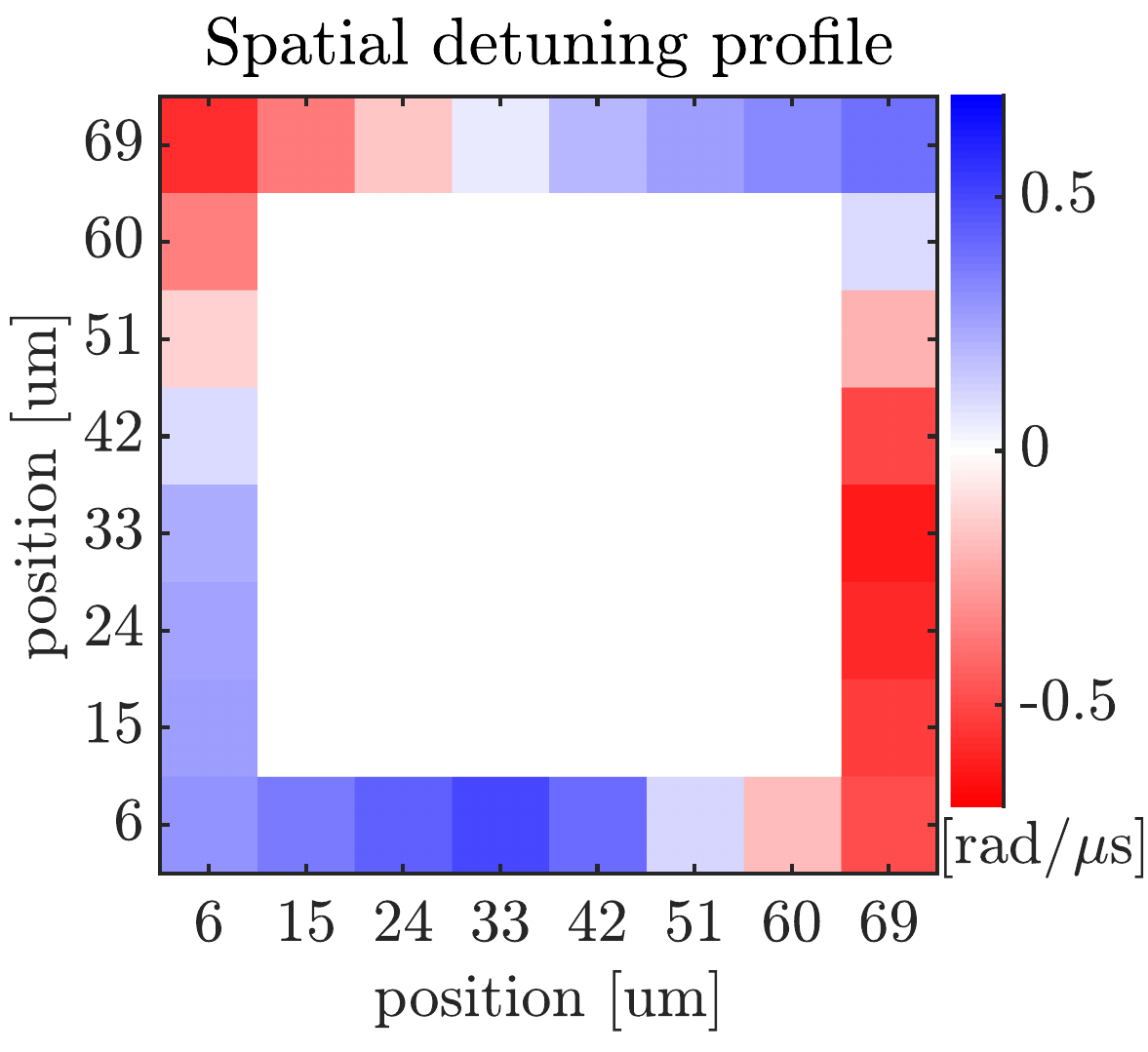}
\caption{Spatial detuning profile on a 28 atom periodic chain. The profile is roughly periodic across the chain, and should be considered as a long-wavelength fluctuation of the detuning on the set global detuning value.}
\label{SMFig4}
\end{figure}

As already explained in End Matter, we find the spatial profile of the detuning fluctuations via bootstrapping each atom individually. We do not find significant changes of this profile over multiple experiments performed over a period of 4 months. Figure~\ref{SMFig4} shows this profile over the many-body chain of 28 atoms with 9 $\mu$m distance between each atom. This profile can be approximated to be periodic, and hence acts like a shallow superlattice potential on the atoms. Our full numerical simulations take this spatial detuning profile into account. 

\section{\label{sec:3}Details of the numerical modeling on tensor networks}

\begin{figure}
\centering
\includegraphics[width=0.9\columnwidth]{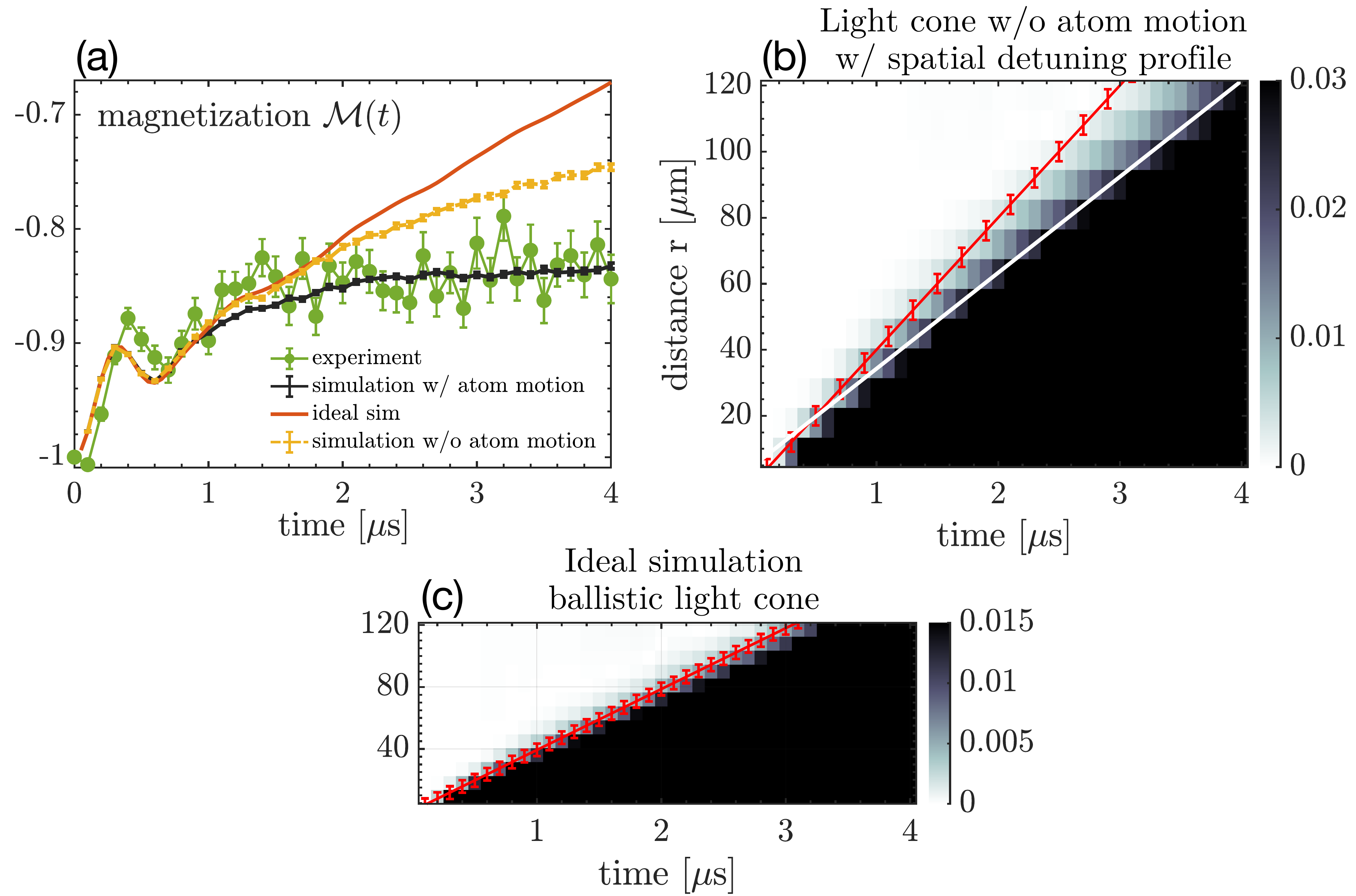}
\caption{Numerical modeling without atom motion. (a) Magnetization with atom motion is black as in Letter, without atom motion, but with spatial detuning profile, is depicted with orange, cannot capture the saturation in later times. (b) Light cone without atom motion but with spatial detuning profile does not exhibit a sub-ballistic behavior, rather the ballistic propagation continues with a renormalized correlation speed (white-solid). (c) Ideal simulation exhibits the theoretically expected ballistic light cone.}
\label{SMFig2}
\end{figure}

We consider atom motion, spatial detuning profile, shot-to-shot fluctuations in the detuning and Rabi frequency in our numerical modeling. When atom motion is not considered in tensor network simulations, the sub-ballistic behavior in the light cones and the saturation of the magnetization over time observed in the experiment cannot be explained. Fig.~\ref{SMFig2} shows the experimental data against the numerical modeling without atom motion. Let us note that the shot-to-shot fluctuations are modeled by normal distribution. Hence it is reasonable to expect their effect on the final results, averaged over many bit strings, to be negligible, which is what we also observe in numerics [not shown]. Spatial detuning profile has a non-negligible effect, and indeed exhibits a slowing down of the magnetization decay, however we still do not observe a saturation as visible in experimental data. More importantly, the light cone without atom motion does not exhibit a sub-ballistic spread of correlations. Rather the propagation of correlations is still ballistic with a renormalized speed, $v_c'=29.1\text{ m}/\rm{s}$. This is straightforward to understand as the spatial detuning profile simply imprints a superlattice potential on the emergent quasi-particles of the many-body system, modifying the correlation speeds rather than the nature of it. 

\begin{figure}
\centering
\includegraphics[width=0.9\columnwidth]{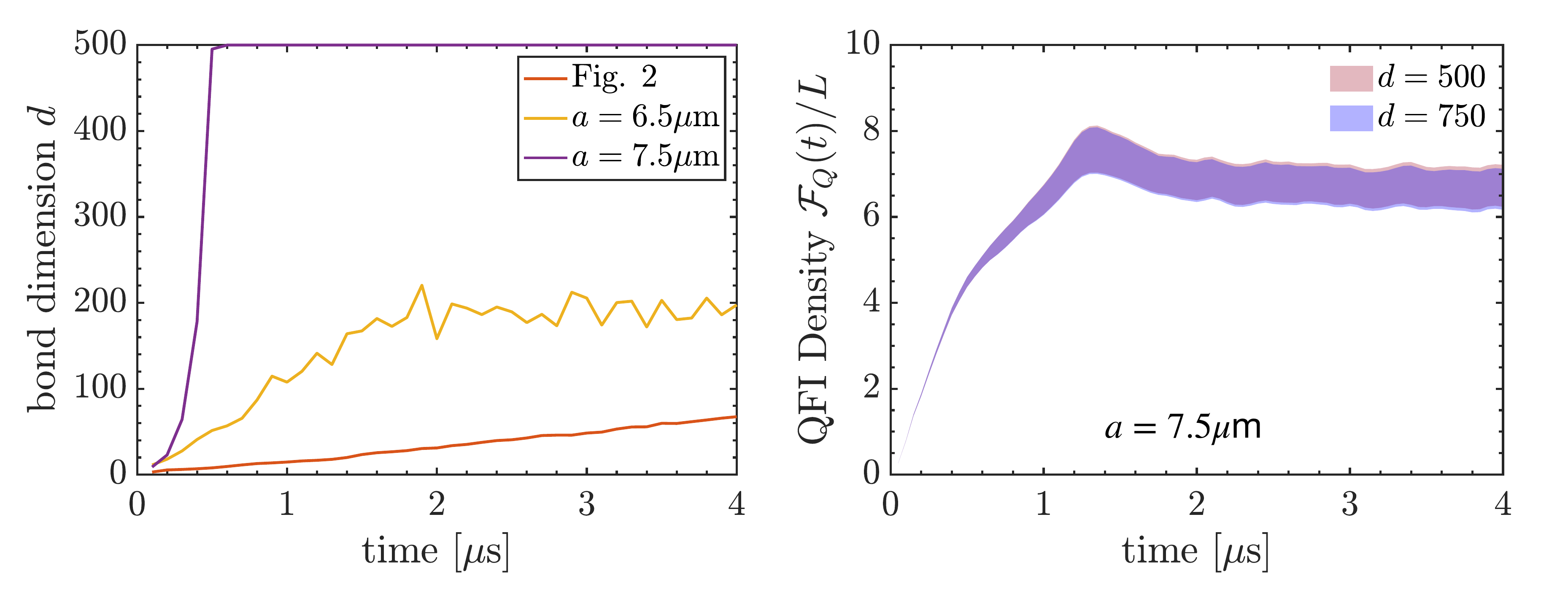}
\caption{(a) The bond dimension over time for different setups. When the quench is performed to the ordered side of the critical point, the maximum bond dimension is typically not saturated. Red: the specs are the same with the Fig.~2 experiment in the main text. Yellow and purple: the specs are the same with the Fig.~3 experiments in the main text, except for different lattice spacings as seen in the legend. (b) QFI Density for a quench at $\Omega=11.7$ rad/$\mu$s and $a=7.5\mu$m (the case depicted in purple in (a)) is shown not to have a significance change as the maximum bond dimension increases from $500$ to $750$.}
\label{SMFig6}
\end{figure}

We set a maximum bond dimension of $500$, a truncation accuracy of $10^{-8}$ and a time increment of $\delta t=0.01\mu$s, and monitor over time how the bond dimension grows for the data in Fig. 2 in the main text, see Fig.~\ref{SMFig6}. Since the bond dimension over time does not reach the maximum bond dimension, we are convinced that the tensor network simulation is fully converged. We find that the reached bond dimension saturates the maximum bond dimension once the quench is done to a point near the critical point and past beyond it, see Fig.~\ref{SMFig6} left panel. This is not surprising, as at these regions the entanglement of the dynamical state builds up more quickly compared to the quenches to the points deep in the ordered phase. We pushed the maximum bond dimension to 750 in this case to check if there is a qualitative change in the measured observables. Fig.~\ref{SMFig6} right panel shows no significant change in the QFI density. Finally, being able to observe a fairly good match between experimental and numerical results is a further evidence that the chosen maximum bond dimension is sufficient for the numerical modeling. 

In our tensor network modeling we encode the exact rectangular lattice together with the next nearest neighbor interactions both at the edges and across the corners. So, the corner effects, although observed to be small, are already included in our numerical modeling.

\section{\label{sec:5}Unitary minimal model}

Let us first derive Eq.~(3) in the main text starting from
\begin{eqnarray}
H(t) &=& \frac{C_6}{4a^6} \sum_r^L \sigma^z_r \sigma^z_{r+1} + \frac{\Omega(t)}{2} \sum_r^L  \sigma^x_r -  \sum_r^L \frac{1}{2}(\Delta(t)-\Delta_{\rm ising})\sigma^z_r, \label{eq1}
\end{eqnarray}
where we omit the next nearest neighbor interactions, and $\Delta_{\rm ising}= \frac{C_6}{a^6}$. To incorporate the positional uncertainty, substitute $a \rightarrow a+\delta_r$ where $\delta_r$ encodes the fluctuation in the distance between two atoms due to atom motion with momentum $p_r$ and atom mass $m$. 
\begin{eqnarray}
H(t) &=& \sum_r^L \frac{C_6}{4(a+\delta_r)^6}  \sigma^z_r \sigma^z_{r+1} + \frac{\Omega(t)}{2} \sum_r^L  \sigma^x_r -  \sum_r^L \frac{1}{2}\bigg(\frac{C_6}{a^6}-\frac{C_6}{(a+\delta_r)^6}\bigg)\sigma^z_r + \sum_r^L \frac{p_r^2}{2m}, \\
&=& \sum_r^L \frac{C_6}{4a^6} \bigg(1+\frac{\delta_r}{a}\bigg)^{-6}  \sigma^z_r \sigma^z_{r+1} + \frac{\Omega(t)}{2} \sum_r^L  \sigma^x_r -  \sum_r^L \frac{1}{2}\bigg(\frac{C_6}{a^6}-\frac{C_6}{a^6}\bigg(1+\frac{\delta_r}{a}\bigg)^{-6}\bigg)\sigma^z_r + \sum_r^L \frac{p_r^2}{2m}, \\
&\approx & \sum_r^L \frac{C_6}{4a^6} \bigg(1-\frac{6\delta_r}{a}\bigg)  \sigma^z_r \sigma^z_{r+1} + \frac{\Omega(t)}{2} \sum_r^L  \sigma^x_r -  \sum_r^L \frac{3\delta_r}{a}\sigma^z_r + \mathcal{O}\bigg(\frac{\delta_r^2}{a^2}\bigg) + \sum_r^L \frac{p_r^2}{2m}. \label{eq4}
\end{eqnarray}
Eq.~\ref{eq4} gives rise to the structure of the normal distributions for the random spin interactions and the longitudinal field. Let us note that the last term, the total atom momenta due to the semiclassical treatment of the thermal fluctuations, is a constant in our calculation because each atom velocity is initially set by the temperature, 15$\mu$K in the tweezer traps, by sampling from the Maxwell-Boltzmann distribution, and we do \textit{not} consider the backaction of the Rydberg interactions on the atom motion, i.e.,~atom velocity for each atom is fixed throughout the time evolution. 

In the semiclassical limit, the spin-motion coupling can be captured by an estimation of the force on the atoms due to the Rydberg interaction, and such a force modifies the classical motion of the atoms by simply inducing acceleration. We estimate the force to be $\approx $ 6.6$\times 10^{-22} \rm kg \cdot m/\rm s^2$ for the specs of the Fig.~2 in the Letter, which corresponds to $\approx 0.005 \rm m/\mu s^2$ acceleration. Such a force only modestly modifies the atom velocity, e.g.,~0.0187 m/s at the end of 4$\mu$s time evolution. This gives rise to roughly $\approx 0.04\mu$m change in atom positions at the end of time evolution. This is almost an order of magnitude smaller than the total effect of the initial positional uncertainty $0.1\mu$m and the atom position attained at $4\mu$s in the absence of acceleration $0.2\mu$m. Although we could expect some effect from the omitted spin-motion coupling, at the same time we do not expect a change in the essence of the physics. This is partly because the minimal model we wrote down based on the emergence of disorder matches well with the atom motion simulations in the absence of the backaction of Rydberg interactions for all parameter sets, and the underlying physics of emerging disorder due to thermal fluctuations will not change upon adding the spin-motion coupling into the picture, albeit higher order effects can be captured and the match with experiment can be improved. 

We also do not expect the spin-motion coupling to be the culprit of the discrepancy between theory and experiment reported for the experiments with different lattice spacings, Fig. 3 in the main text. This is because the force due to Rydberg interaction will be the strongest in the smallest lattice spacing, $a=6\mu$m, however this is not where we observe the largest discrepancy.  Furthermore the disorder strength is fitted to be even smaller than $\delta_r=0.1\mu$m within the unitary minimal model where $\delta_r$ is treated as a parameter to be determined by the experiment. This is in stark contrast to the expected effect of atom acceleration at short distances.

Incorporating the force due to Rydberg interaction into the model requires a full quantum formalism exactly due to the spin-motion coupling. This could be done by promoting fluctuations in the atomic distance between phonon operators $\delta_r \rightarrow \frac{1}{\sqrt{2m\omega}}\left(\hat b_{r+1} + \hat b^{\dagger}_{r+1} - \hat b_r - \hat b^{\dagger}_r\right)$.
\begin{eqnarray}
\mathcal{H} &=& \sum_r^L \frac{C_6}{4a^6} \bigg(1-\frac{6}{a} \frac{1}{\sqrt{2m\omega}}\left(\hat b_{r+1} + \hat b^{\dagger}_{r+1} - \hat b_r - \hat b^{\dagger}_r\right) \bigg)  \sigma^z_r \sigma^z_{r+1} + \frac{\Omega(t)}{2} \sum_r^L  \sigma^x_r \notag \\
&-&  \sum_r^L \frac{3}{a} \frac{1}{\sqrt{2m\omega}}\left(\hat b_{r+1} + \hat b^{\dagger}_{r+1} - \hat b_r - \hat b^{\dagger}_r\right)\sigma^z_r -\frac{\omega}{4} \sum_r^L \left(\hat b^{\dagger}_r-\hat b_r\right)^2. \label{eq:5}
\end{eqnarray}
While we made an estimation above in the semiclassical limit, the degree of spin-motion entanglement in the many-body wave function can be accurately determined from this quantized Hamiltonian. However this requires the full solution of this model, and is an interesting question to answer in future research. 

Let us also note that to see disorder emerging even in this quantized Hamiltonian, one does not need spin-motion entanglement. By considering a local phononic field $\phi_p^r$ in the mean field limit, which is a reasonable assumption to capture the boson coupling as an effective disorder effect on the atomic array, we will end up with $\langle \phi_p^r \rangle = \bigg \langle \left(\hat b_{r+1} + \hat b^{\dagger}_{r+1} - \hat b_r - \hat b^{\dagger}_r\right) \bigg \rangle$. To obtain $\langle \phi_p^r \rangle \neq 0$ leading to the emergence of disorder from microscopics, one needs the phononic state to be a distribution over the Fock basis, e.g.,~coherent, thermal or coherent-thermal states. This already tells us that in our experimental setup the atoms are not occupying the vibrational ground state due to finite temperature fluctuations, which is consistent with the experimental observations. Once $\langle \phi_p^r \rangle \neq 0$ is achieved, the disorder strength would follow from the microscopics as $\mathcal{W}=\langle \phi_p^r \rangle /a$. 

Let us note finally, the phonon field being in a single Fock state will give rise to $\langle \phi_p^r \rangle = 0$. This simple argument immediately shows the distinction of our setup from the one theoretically studied in Ref.~\cite{PhysRevLett.132.133401}, where the spin-motion coupling is fully considered within a Hamiltonian similar to Eq.~\ref{eq:5}, and they set the phononic field in the vibrational ground state $\ket{0}$ or in a Fock state with one phonon excitation $\ket{1}$. When the phononic field is treated in mean-field level and hence under the assumption of no spin-motion entanglement, disorder is not expected to emerge from the setup in Ref.~\cite{PhysRevLett.132.133401}. Hence, the sub-ballistic lightcone behavior observed in Ref.~\cite{PhysRevLett.132.133401} when the phonon state was prepared in $\ket{1}$, must originate from a physics different from the disorder. Therefore, the possible localization phenomena that can be observed in Rydberg atom arrays \cite{bernien2017probing,PhysRevLett.132.133401} seem to originate from different sources, demonstrating the richness of the physics of this platform while differentiating the emergence of disorder we found in this work from the literature.

\subsection{Numerics on the minimal model}

\begin{figure}
\centering
\includegraphics[width=0.8\columnwidth]{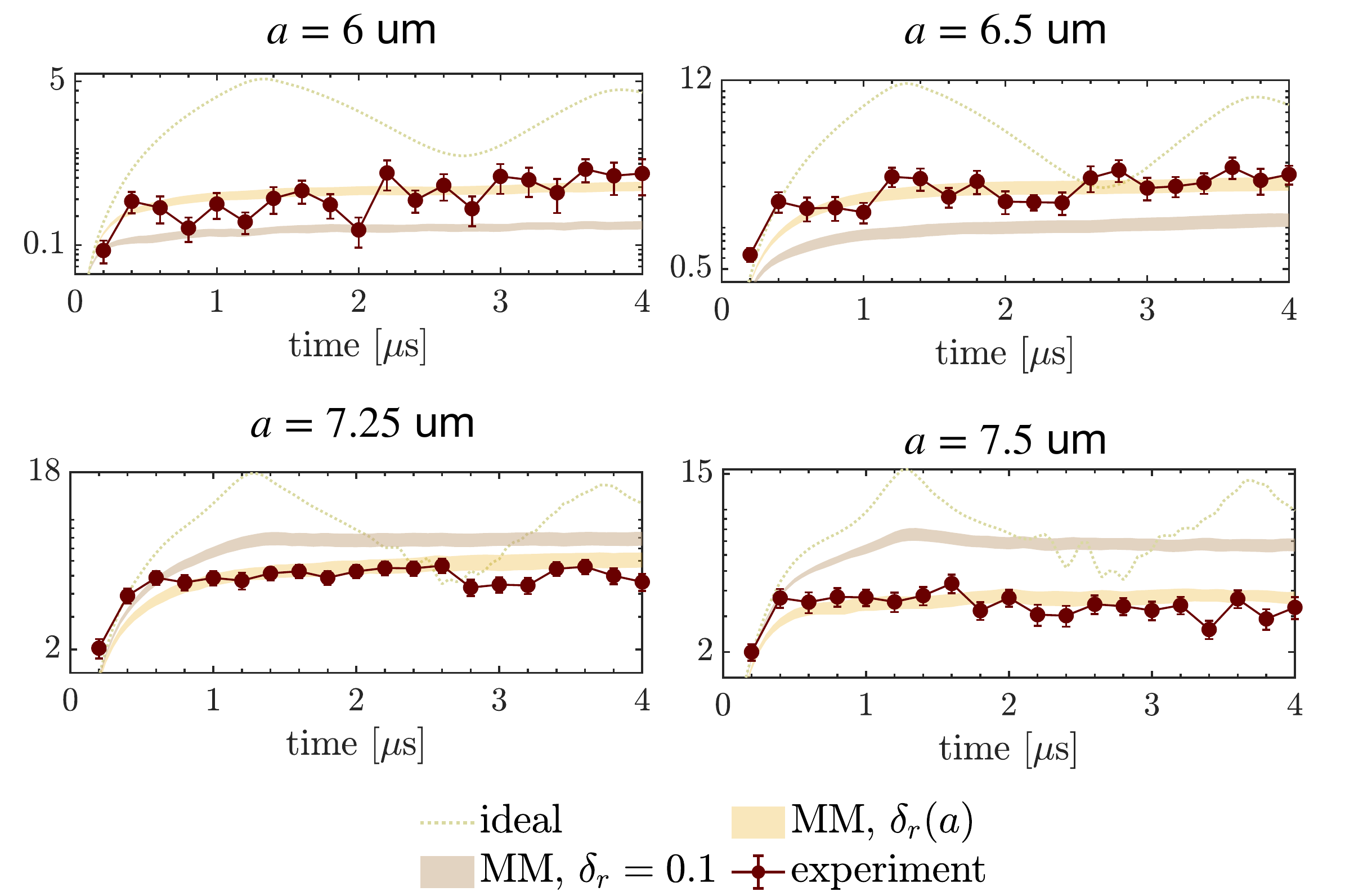}
\caption{Quantum Fisher information for lattice spacings $a=6,6.5,7.25,7.5\mu$m showing the experiment data, ideal simulation, MM simulations with fixed positional uncertainty $\delta_r=0.1\mu$m and the fitted positional uncertainty with respect to lattice distance $\delta_r(a)$. The resulting $\delta_r(a)$ determines the disorder strength of the unitary MM.}
\label{SMFig3}
\end{figure}
In this section, we share additional numerical results. (1) The unitary minimal model with fitted positional uncertainty $\delta_r(a)$ at different lattice constants $a$ is a good theoretical model that matches the experimental results, Fig.~\ref{SMFig3}. The resulting disorder profile can be utilized to further study the unitary MM. 

\begin{figure}
\centering
\includegraphics[width=0.8\columnwidth]{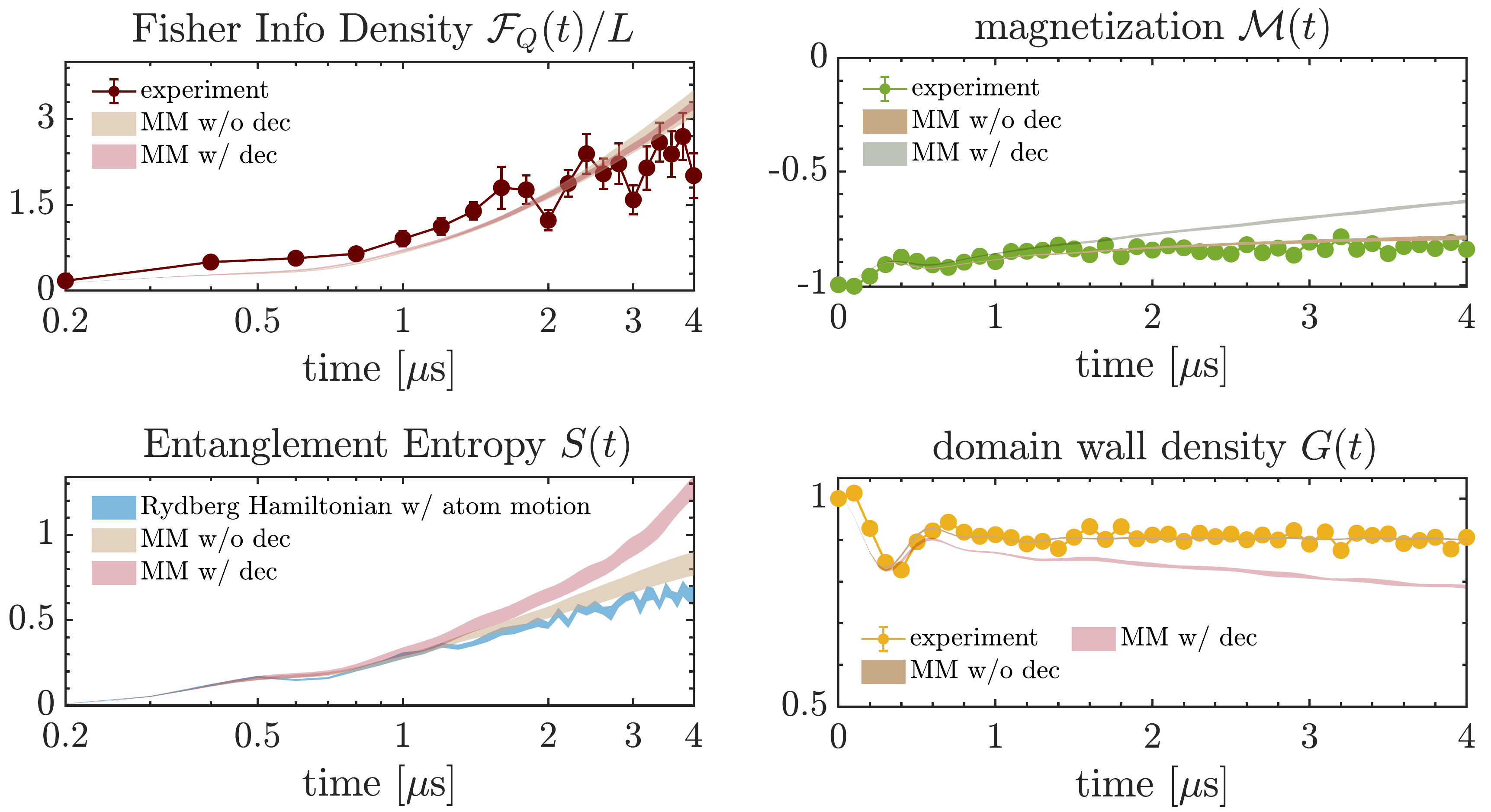}
\caption{The minimal model with and without decoherence in Fisher information density, entanglement entropy, magnetization and domain-wall density. All plots compare the results of these numerical calculations to the experiment, except in the entanglement entropy, where we compare them to the Rydberg atom array Hamiltonian simulation with the atom motion treated classically.}
\label{SMFig8}
\end{figure}

(2) The (unitary) minimal model at the expected positional uncertainty $\delta_r=0.1\mu$m reproduces the experimental results presented in Fig.~(2) in the main text, see Fig.~\ref{SMFig8}. Here the experimental data was collected at $a=9\mu$m and $\Omega=2.21$ rad/$\mu$s. We also observe in Fig.~\ref{SMFig8} that adding the fitted decoherence level to the MM does not have an effect on the QFI density. However, decoherence slightly changes the magnetization,  domain-wall density, and entanglement entropy. In fact, the behaviors of atom motion and decoherence on the many-body dynamics are distinct in these observables: atom motion stabilizes magnetization and domain-wall density, whereas decoherence causes a tendency in them to decay. For entanglement entropy, atom motion leads to logarithmic increase in time, whereas adding decoherence leads to linear increase in time. It is not unusual to see the effect of the same decoherence in different observables differently, as some observables are more susceptible to the decoherence effects. All in all, we conclude that the experimental data in Fig.~(2) in the main text are not significantly affected by the decoherence processes in the experimental setup.

\begin{figure}
\centering
\includegraphics[width=0.8\columnwidth]{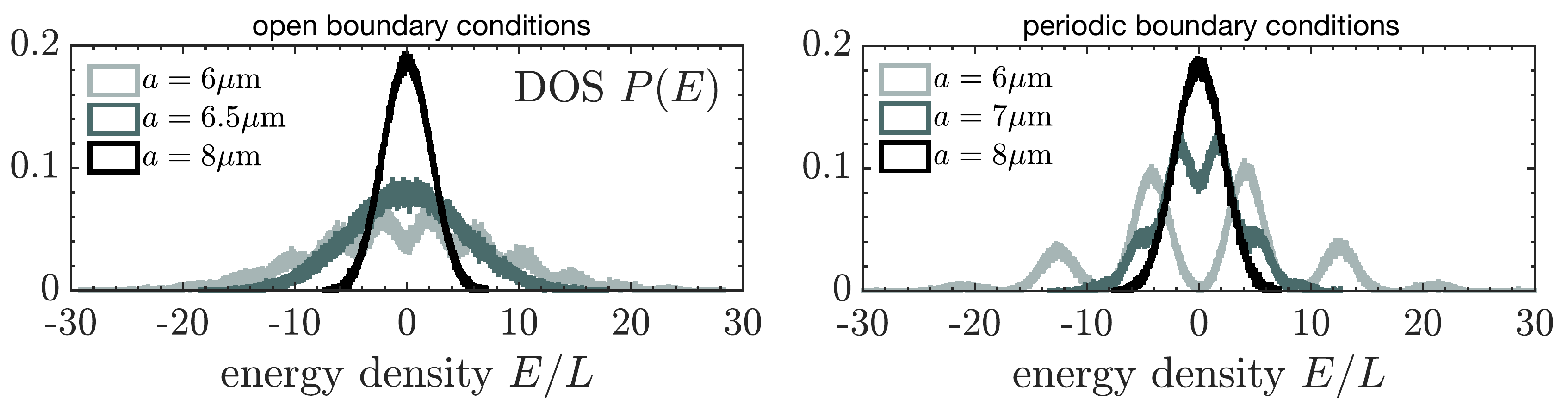}
\caption{Density of states in the minimal model at different lattice distances for open (left panel) and periodic (right panel) boundary conditions.}
\label{SMFig7}
\end{figure}

(3) We diagonalize the unitary minimal model (MM) Hamiltonian with disorder strengths $\mathcal{W}$  determined based on the fitted positional uncertainty for $L=14$, and examine the density of states (DoS) at different lattice constants and boundary conditions in Fig.~\ref{SMFig7}. In both boundary conditions, mini-bands appear in smaller lattice constants, while the mini-bands are more clear in the periodic boundary conditions. Increasing the lattice constants where we have chaotic spectral statistics with $\langle r \rangle=0.53$, where $\langle r \rangle = \langle \text{min}[E_{n+1}-E_{n},E_n-E_{n-1}]/\text{max}[E_{n+1}-E_{n},E_n-E_{n-1}] \rangle_n$ where $E_n$ is the $n^{\text{th}}$ eigenenergy \cite{PhysRevB.75.155111}, the DoS becomes Gaussian.

\section{\label{sec:6}Ideal time-averaged observables in experimental time interval}

\begin{figure}
\centering
\includegraphics[width=0.8\columnwidth]{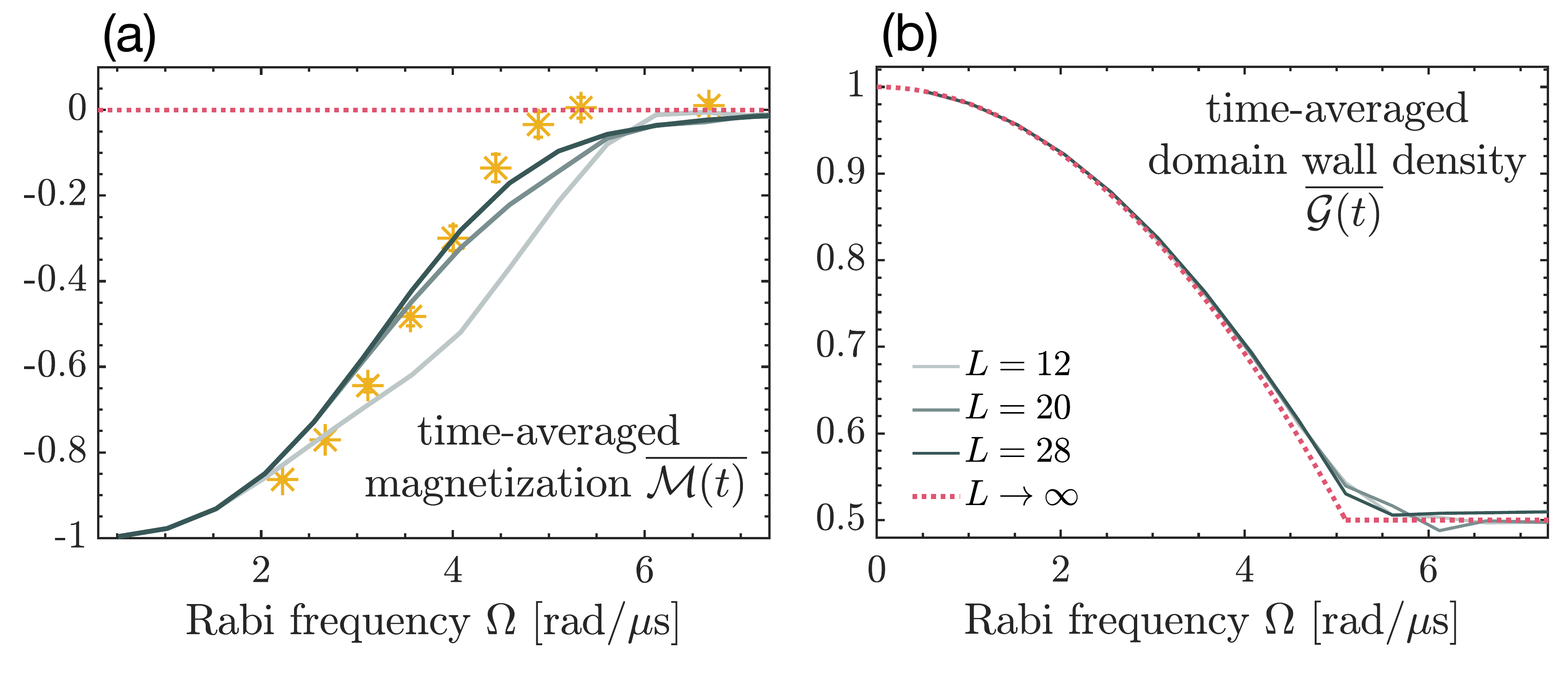}
\caption{The time-averaged (a) magnetization and (b) domain wall density in ideal conditions for three different system sizes $L=12,20,28$. The pink dotted line shows the results in the thermodynamic limit and at infinite time. Subfigure (a) includes the full simulation with uncertainties at $L=28$ for comparison.}
\label{SMFig5}
\end{figure}

Here we share the simulations of time-averaged magnetization and domain-wall density averaged between $[0,4]\mu$s omitting the atom motion, hence in ideal conditions, Figure~\ref{SMFig5}. The tensor network simulations are still performed for the Rydberg atom array, and they take into account the next-nearest neighbor interactions. Note that these observables in the clean TFIM exhibit strong finite-size effects in their time evolution as revivals, as already seen in Figure 3(b$_1$) in the main text. Hence, the time average over $[0,4]\mu$s actually also averages over such finite-size effects.

Specifically we observe the tendency of magnetization to smoothen around the critical point, $\Omega \sim 5 \text{ rad/}\mu$s as the system size increases, in contrast to the experimental data (see main text) and the full simulation with atom motion (yellow-stars), which sharpen around the critical point. This is expected, as the infinite-time and thermodynamic limit response in ideal conditions is featureless, the pink-dotted line in (a). Hence, we observe once more the important role of atom motion in modeling the experiment and capturing the correct physical interpretation behind the experimental data. Fig.~\ref{SMFig5}(b) shows the tendency of the domain-wall density to sharpen around the critical point as the system size increases in the ideal conditions. This is a reason why this quantity was proposed to probe the quantum critical point of TFIM in the first place \cite{PhysRevLett.123.115701}. However as already shown in the Letter, Figure~3, the longitudinal field generated by the atom motion at the TFIM limit causes the domain-wall density to undershoot $\overline{\mathcal{G}(t)} = 1/2$. Experimental data also looks smoother compared to the data in ideal conditions, which we attribute to the presence of significant atom motion.

\section{\label{sec:2}Error mitigation technique}

Perhaps the most important uncertainty in the Aquila Rydberg atom simulator is the readout errors of the final bit strings. Let us define $p_{i\rightarrow j}$ as the probability of state $\ket{i}$ to be measured as state $\ket{j}$ where $i,j=0,1$. Naturally, $p_{i\rightarrow i}+p_{i\rightarrow j}=1$ for $j\neq i$. For Aquila, $p_{0\rightarrow 1}=0.01$ and $p_{1\rightarrow 0}=0.05$.

A confusion matrix $\mathcal{C}$ encodes all readout errors. For example, $\mathcal{C}$ for one-site measurements follows as,
\begin{eqnarray}
\mathcal{C} &=& \left(
    \begin{matrix}
    p_{0\rightarrow 0} & p_{1\rightarrow 0} \\
    p_{0\rightarrow 1} & p_{1\rightarrow 1} 
    \end{matrix} \right).
\end{eqnarray}
Assuming the one-site measurement is in $Z$ basis, the error-mitigated expectation value of single-site magnetization $   \braket{\tilde{\sigma}_i^z}  $ at any time and site will be,
\begin{eqnarray}
   \braket{\tilde{\sigma}_i^z} &=& \frac{1}{N} \sum_{k=1}^N \bigg( -\bra{0} \mathcal{C}^{-1} \ket{s_k} + \bra{1} \mathcal{C}^{-1} \ket{s_k}  \bigg),
\end{eqnarray}
where $\ket{0}=(1, 0)^T$ and $\ket{1}=(0, 1)^T$, and $\ket{s_k}$ denotes either of two states that was measured in the experiment in the $k^{\rm th}$ shot. Summation over $k$ is the statistical average over $N$ many shots. 

We also measure two-site observables such as $\braket{\sigma_z^i(t) \sigma_z^{j}(t)}$. Confusion matrix for such observables are written as,
\begin{eqnarray}
\mathcal{C} &=& \left(
    \begin{matrix}
    p_{00 \rightarrow 00} & p_{01 \rightarrow 00} & p_{10 \rightarrow 00} &  p_{11 \rightarrow 00} \\
   p_{00 \rightarrow 01} & p_{01 \rightarrow 01} & p_{10 \rightarrow 01} & p_{11 \rightarrow 01}\\
   p_{00 \rightarrow 10} & p_{01 \rightarrow 10} & p_{10 \rightarrow 10} & p_{11 \rightarrow 10} \\
   p_{00 \rightarrow 11} & p_{01 \rightarrow 11} &  p_{10 \rightarrow 11} & p_{11 \rightarrow 11}
    \end{matrix} \right),
\end{eqnarray}
where the following holds with $p_{ij \rightarrow kl}$ denoting two qubit readout error probabilities,
\begin{eqnarray}
 p_{00 \rightarrow 01}&=&p_{00 \rightarrow 10} = p_{0\rightarrow 1}  (1-p_{0\rightarrow 1}).  \\
 p_{00 \rightarrow 11} &=& p_{0\rightarrow 1}^2 \notag \\
p_{00 \rightarrow 00} &=&  (1-p_{0\rightarrow 1})^2.\notag \\
p_{11 \rightarrow 01}&=&p_{11 \rightarrow 10} = p_{1\rightarrow 0}  (1-p_{1\rightarrow 0}). \notag \\
 p_{11 \rightarrow 00} &=& p_{1\rightarrow 0}^2 \notag\\
p_{11 \rightarrow 11} &=&  (1-p_{1\rightarrow 0})^2.\notag \\
p_{01 \rightarrow 01} &=& (1-p_{0\rightarrow 1}) (1-p_{1\rightarrow 0}) \notag\\
p_{01 \rightarrow 00} &=& (1-p_{0\rightarrow 1}) p_{1\rightarrow 0}\notag \\
p_{01 \rightarrow 11} &=&  p_{0\rightarrow 1} (1-p_{1\rightarrow 0})\notag \\
p_{01 \rightarrow 10} &=& p_{0\rightarrow 1} p_{1\rightarrow 0} \notag\\
p_{10 \rightarrow 10} &=& (1-p_{1\rightarrow 0}) (1-p_{0\rightarrow 1}) \notag\\
p_{10 \rightarrow 11} &=& (1-p_{1\rightarrow 0}) p_{0\rightarrow 1}\notag \\
p_{10 \rightarrow 00} &=&  p_{1\rightarrow 0} (1-p_{0\rightarrow 1})\notag \\
p_{10 \rightarrow 01} &=& p_{1\rightarrow 0} p_{0\rightarrow 1} \notag
\end{eqnarray}
This leads to the error-mitigated two-point correlator $\braket{\widetilde{\sigma_z^i\sigma_z^j}}$
\begin{eqnarray}
  \braket{\widetilde{\sigma_z^i \sigma_z^j}} &=& \frac{1}{N} \sum_{k=1}^N \bigg( \bra{00}+ \bra{11} - \bra{01} -\bra{10}\bigg) \mathcal{C}^{-1} \ket{s_k s'_k}.\notag
\end{eqnarray}

While these can be numerically calculated for each many-body bit string, due to the simplicity of the confusion matrix we can write analytical formulae for magnetization $\mathcal{M}$ and nearest-neighbor two-point correlator $\mathcal{G}$. Consider a general $n$-qubit state
\begin{eqnarray}
\ket{\psi}=\sum_{m=1}^{2^n}A_m \ket{S_m}
\end{eqnarray}
where $\ket{S_m}$ are $2^n$ basis states. Each of the states corresponds to a bit string such as $\ket{0110...10}$. Let
\begin{eqnarray}
\ket{S_m}=\ket{s_1^{(m)}}\ket{s_2^{(m)}}...\ket{s_k^{(m)}}...\ket{s_n^{(m)}}.
\end{eqnarray}
The error-free magnetization $\widetilde{\cal{M}}$ is
\begin{eqnarray}
\widetilde{\cal{M}}=\frac{1}{n}\sum_{k=1}^n\braket{\sigma_k^z}=\frac{1}{n}\sum_{k=1}^n \sum_{m=1}^{2^n}|A_m|^2\braket{S_m|\sigma_k^z|S_m}=\frac{1}{n}\sum_{k=1}^n \sum_{m=1}^{2^n}|A_m|^2\braket{s_k^{(m)}|\sigma_k^z|s_k^{(m)}}
\end{eqnarray}
with
\begin{eqnarray}
\braket{s_k^{(m)}|\sigma_k^z|s_k^{(m)}}=
\begin{cases}
+1 &,\,\ket{s_k^{(m)}}=\ket{1}\\
-1 &,\,\ket{s_k^{(m)}}=\ket{0}
\end{cases}
\end{eqnarray}
For any site $k$, there is a probability $p_{i\rightarrow j}$ to replace $\braket{i|\sigma_k^z|i}$ by $\braket{j|\sigma_k^z|j}$. The magnetization with readout errors is then
\begin{eqnarray}
\cal{M}&=&\frac{1}{n}\sum_{k=1}^n\bigg\{\sum_{m\in\{m'|s_k^{(m')}=1\}}\!\!\!\!\!\!\!|A_m|^2\braket{s_k^{(m)}|\sigma_k^z|s_k^{(m)}}+\sum_{m\in\{m'|s_k^{(m')}=0\}}\!\!\!\!\!\!\!|A_m|^2\braket{s_k^{(m)}|\sigma_k^z|s_k^{(m)}}\bigg\}\\
&=&\frac{1}{n}\sum_{k=1}^n\bigg\{\sum_{m\in\{m'|s_k^{(m')}=1\}}\!\!\!\!\!\!\!|A_m|^2\bigg[p_{1\rightarrow 1}(+1)+p_{1\rightarrow 0}(-1)\bigg]+\!\!\!\!\!\sum_{m\in\{m'|s_k^{(m')}=0\}}\!\!\!\!\!\!\!|A_m|^2\bigg[p_{0\rightarrow 0}(-1)+p_{0\rightarrow 1}(+1)\bigg]\bigg\} \notag \\
&=&\frac{1}{n}\sum_{k=1}^n\bigg\{a_k
(p_{1\rightarrow 1}-p_{1\rightarrow 0})+b_k(p_{0\rightarrow 1}-p_{0\rightarrow 0})\bigg\}\notag
\end{eqnarray}
where we have defined $a_k=\sum_{m\in\{m'|s_k^{(m')}=1\}}|A_m|^2$ and $b_k=\sum_{m\in\{m'|s_k^{(m')}=0\}}|A_m|^2$, with $a_k+b_k=1$. Using these quantities, the error-free magnetization can be written as
\begin{eqnarray}
\widetilde{\cal{M}}=\frac{1}{n}\sum_{k=1}^n(a_k-b_k)
\end{eqnarray}
Defining $a=\frac{1}{n}\sum_{k=1}^n a_k$ and $b=\frac{1}{n}\sum_{k=1}^n b_k$, we obtain the following set of equations
\begin{eqnarray}
\widetilde{\cal{M}}&=&a-b\\
\cal{M}&=&a(p_{1\rightarrow 1}-p_{1\rightarrow 0})+b(p_{0\rightarrow 1}-p_{0\rightarrow 0}) \notag\\
1&=&a+b \notag
\end{eqnarray}
Therefore,
\begin{eqnarray}
\widetilde{\mathcal{M}}=2\frac{\mathcal{M}+p_{0\rightarrow 0}-p_{0\rightarrow 1}}{p_{1\rightarrow 1}-p_{1\rightarrow 0}-p_{0\rightarrow 1}+p_{0\rightarrow 0}}-1.
\end{eqnarray}

Aquila has $p_{0\rightarrow 0} = 0.99$, $p_{0\rightarrow 1}=0.01$, $p_{1\rightarrow 0}=0.05$ and $p_{1\rightarrow 1}=0.95$. For the error-free nearest-neighbor two-point correlator
\begin{eqnarray}
\widetilde{\cal{G}}&=&\frac{1}{n}\sum_{k=1}^n\braket{\sigma_k^z \sigma_{k+1}^z}=\frac{1}{n}\sum_{k=1}^n \sum_{m=1}^{2^n}|A_m|^2\braket{s_k^{(m)} s_{k+1}^{(m)}|\sigma_k^z \sigma_{k+1}^z|s_k^{(m)}s_{k+1}^{(m)}}\\
&=&\frac{1}{n}\sum_{k=1}^n(c_k-d_k-e_k+f_k) \notag\\
&=&c-d-e+f \notag
\end{eqnarray}
where we have defined
\begin{eqnarray}
c_k&=& \sum_{m\in\{m'|s_k^{(m')}=0,s_{k+1}^{(m')}=0\}}\!\!\!\!\!\!\!\! |A_m|^2\\
d_k&=& \sum_{m\in\{m'|s_k^{(m')}=0,s_{k+1}^{(m')}=1\}}\!\!\!\!\!\!\!\!|A_m|^2 \notag\\
e_k&=& \sum_{m\in\{m'|s_k^{(m')}=1,s_{k+1}^{(m')}=0\}}\!\!\!\!\!\!\!\!|A_m|^2 \notag\\
f_k&=& \sum_{m\in\{m'|s_k^{(m')}=1,s_{k+1}^{(m')}=1\}}\!\!\!\!\!\!\!\!|A_m|^2 \notag
\end{eqnarray}
and $c=\frac{1}{n}\sum_{k=1}^n c_k$, $d=\frac{1}{n}\sum_{k=1}^n d_k$, $e=\frac{1}{n}\sum_{k=1}^n e_k$, $f=\frac{1}{n}\sum_{k=1}^n f_k$. The two-point correlator with the readout errors is
\begin{eqnarray}
\cal{G} &=& c\bigg[p_{00 \rightarrow 00}(+1)+p_{00 \rightarrow 01}(-1)+p_{00 \rightarrow 10}(-1)+p_{00 \rightarrow 11}(+1)\bigg]\\
&&+d\bigg[p_{01 \rightarrow 00}(+1)+p_{01 \rightarrow 01}(-1)+p_{01 \rightarrow 10}(-1)+p_{01 \rightarrow 11}(+1)\bigg] \notag\\
&&+e\bigg[p_{10 \rightarrow 00}(+1)+p_{10 \rightarrow 01}(-1)+p_{10 \rightarrow 10}(-1)+p_{10 \rightarrow 11}(+1)\bigg]\notag\\
&&+f\bigg[p_{11 \rightarrow 00}(+1)+p_{11 \rightarrow 01}(-1)+p_{11 \rightarrow 10}(-1)+p_{11 \rightarrow 11}(+1)\bigg] \equiv cx+dy+ez+fw. \notag
\end{eqnarray}
Notice that $y=z$. One can also express the error-free magnetization from these quantities
\begin{eqnarray}
\widetilde{\cal{M}}=\frac{1}{2}(2f-2c+0\cdot d+0\cdot e)=f-c
\end{eqnarray}
leading to the following set of equations
\begin{eqnarray}
\widetilde{\cal{G}}&=&c-d-e+f\\
\cal{G}&=&cx+dy+ey+fw \notag\\
\widetilde{\cal{M}}&=&f-c \notag\\
1&=&c+d+e+f \notag
\end{eqnarray}
Therefore,
\begin{eqnarray}
\widetilde{\cal{G}}=-\frac{w+x+2y+2(w-x)\widetilde{\cal{M}}-4\cal{G}}{w+x-2y}. 
\end{eqnarray}

\end{document}